\documentclass[fleqn,usenatbib]{mnras}
\usepackage[T1]{fontenc}
\usepackage{ae,aecompl}

\usepackage{graphicx}	
\usepackage{amsmath}	
\usepackage{amssymb}	
\usepackage{tikz}
\usetikzlibrary[patterns]
\usetikzlibrary{decorations.pathreplacing}
\usepackage{tcolorbox}
\usepackage{courier}

\title[Inferring Distribution Functions]{Inferring Time-Dependent Distribution Functions from Kinematic Snapshots}

\author[Keir Darling et al.]{
	Keir Darling\thanks{E-mail: keir.darling@queensu.ca}
	and Lawrence M. Widrow\thanks{E-mail: widrow@queensu.ca}
	\\
	Department of Physics, Engineering Physics \& Astronomy, Queen's University, Stirling Hall, Kingston, ON K7L 3N6, Canada
}

\date{Accepted XXX. Received YYY; in original form ZZZ}

\pubyear{2021}

\begin{document}
	
	\label{firstpage}
	\pagerange{\pageref{firstpage}--\pageref{lastpage}}
	\maketitle
	
	\begin{abstract}
		We propose a method for constructing the time-dependent phase space distribution function (DF) of a collisionless system from an isolated kinematic snapshot. In general, the problem of mapping a single snapshot to a time-dependent function is intractable. Here we assume a finite series representation of the DF, constructed from the spectrum of the system's Koopman operator. This reduces the original problem to one of mapping a kinematic snapshot to a discrete spectrum rather than to a time-dependent function. We implement this mapping with a convolutional neural network (CNN). The method is demonstrated on two example models: the quantum simple harmonic oscillator, and a self-gravitating isothermal plane. The latter system exhibits phase space spiral structure similar to that observed in Gaia Data Release 2.     
	\end{abstract}
	
	
	
	\section{Introduction}
	
	Disc galaxies, such as the Milky Way, often have bars, spiral arms, and
	warps. Such features are often too large to be treated as linear
	perturbations from an equilibrium model. In this work we introduce a
	novel scheme to construct galaxy models that include
	nonlinear time-dependent phenomena. The method relaxes the common
	assumption that isolated galaxies are close to equilibrium while
	retaining the attractive feature of superposition from perturbation theory. The ultimate goal of this endeavor is to determine a mapping from a kinematic snapshot to a dynamical model. With the Milky Way, the motivation for considering this problem comes from the vast catalog of stellar positions and velocities  being made available by the Gaia mission \citep{gaia2018}. In this case, the dynamical model is described by the stellar phase space distribution function (DF).
	
	In our scheme, the map from kinematic snapshot to DF takes the form of a convolutional neural network (CNN), which we train on a set of N-body simulations. For each simulation, we construct a data matrix where the columns correspond to a time-dependent scalar observable of the phase space coordinates. The observable here is an approximation of the DF on a grid in phase space, that is, the coarse-grained DF.  Our scheme is based on the idea that there exists a {\it linear} operator, that determines the time evolution of observables, even if the evolution in the coordinates is nonlinear. Such an operator is frequently called the Koopman operator \citep{kutz2016,mezic2005,rowleyMezic2009}. It follows that the DF can be expanded as a series using the eigendecomposition of this operator. We approximate the eigenfunctions and eigenvalues of the Koopman operator using a data-driven method known as Dynamic Mode Decomposition (DMD). The CNN then maps snapshots to eigenfunctions and eigenvalues, from which the time-dependent DF is recovered.
	
	The opportunity to infer dynamics from a single kinematic snapshot has been recognized for some time.  For example, \cite{bovyHogg2010} solve the problem of modeling the dynamics of planets in the solar system and the gravitational force law from the positions and velocities of the planets at a single instant in time. Their analysis employs Bayesian inference with the key assumption that the planets are on stable orbits. The steady-state assumption is also at the heart of orbit based schemes such as the Schwarzschild Method \citep{schwarzschild1979} and Made-to-Measure \citep{syer1996}. These approaches build the DF from a set of orbits subject to the condition that the system is in equilibrium and that observables derived from the DF fit observational constraints. More recently, \cite{green2020} proposed Hamiltonian neural networks as a means of inferring the gravitational potential of a collisionless system from a snapshot of the DF.  This approach also assumes a stationary DF in a time-independent potential.
	
	We focus on the collisionless Boltzmann equation (CBE) coupled to gravity through the Poisson equation. The standard analytic method for treating this nonlinear system is to first construct an equilibrium model and then study departures from equilibrium by solving the corresponding linearized equations \citep{BT2008, binney2013}. Conversely, one can simulate the full nonlinear dynamics of galaxies via N-body techniques. One way to connect simulations with analytic methods is via the spectral analysis methods of \cite{sellwood1986}. This approach was originally developed to study the emergence of bars and spiral structure in isolated disc galaxies from equilibrium initial conditions, though it has been extended to include bending waves and warps by \cite{chequers2017}. The idea is to take a sequence of snapshots, each of which comprises a set of observables such as the surface density across the disc, $\Sigma(R,\phi,t)$, where $R$ and $\phi$ are polar coordinates in a disc-centered system. One then computes a Fourier series in azimuthal mode number $m$ and frequency $\omega$: $\tilde{\Sigma}(R,m,\omega)$. Though the method does not explicitly assume that the Fourier components are linear, it is motivated by perturbation theory since linear perturbations about an asymmetric equilibrium state have definite $m$ and parity about the mid-plane. The framework therefore makes it difficult to study nonlinear couplings that may arise with spiral structure and warps \citep{MT1997a, MT1997b}.
	
	The phase space spirals discovered by \cite{antoja2018} in kinematic data from the Gaia mission \citep{gaia2018} provide another example of a galactic structure that is difficult to study using perturbation theory. The spirals appear in the $z-v_z$ projection of the stellar DF where $z$ is the position of a star relative to the mid-plane of the Galaxy and $v_z$ is the the $z$-component of its velocity. This observation is suggestive of phase mixing in $z-v_z$ that follows a perturbation to the disc $300-900\,{\rm Myr}$ ago \citep{antoja2018, binney2018, darling2019,blandhawthorn2019, shen2019, laporte2019}. Mode analysis via perturbation theory has not reportedly reproduced such spiral structure. \cite{mathur1990}, \cite{weinberg1991}, and \cite{widrow2015} determined modes of the isothermal plane, a model for the vertical structure of galactic discs devised by \cite{spitzer1942, camm1950}. These modes included a seiche mode where the slab as a whole oscillates in the gravitational potential of a background halo, as well as higher order modes that correspond to compression-rarefaction or breathing of the disc. \cite{weinberg1991} carried out N-body simulations by passing a perturbing mass through the slab. While the seiche mode was easily excited, the higher order modes were not. A likely explanation for this is that the discrete higher order modes lie close to a continuum and therefore any disturbance tends to rapidly phase-mix away. A more promising use of perturbation theory can be found in \cite{bennetbovy}, in which the authors determine the secular response of a collisionless slab to an external perturbation by solving the linearized CBE.
	
	We propose an alternative means for dealing with nonlinearity in the CBE-Poisson system, which draws on perturbation theory and spectral analysis while addressing their shortcomings. As in spectral analysis, we consider a sequence of snapshots from the evolution of a system. Here we take the snapshots to be samples of the course-grained DF, which acts as a scalar observable function of the phase space of the system. The crux of our method is that the dynamics in the space of observables is linear even if the dynamics in phase space are nonlinear. The basis for the transformation between these two spaces comes from Koopman theory \citep{koopman,mezic2005,rowleyMezic2009}, and the linear Koopman operator that determines the linear dynamics can be estimated from simulations using DMD \citep{schmidt2010,Tu2013}. The advantage of this over the spectral analysis methods of \cite{sellwood1986} is that it does not make {\it a priori} assumptions about the form of the modes. In particular, it does not assume definite $m$ or parity about the mid-plane. Instead, the form of the modes is determined by the data.  Moreover, the equilibrium (or dominant zero-frequency) state and the time-dependent modes are determined simultaneously. A detailed study of the isothermal plane using DMD found that the technique could identify structures leading to long-lived spirals \citep{darling2018} similar to those seen in Gaia data \citep{antoja2018}. In addition, a preliminary study of a disc galaxy simulation produced DMD modes that include both $m=2$ spirals and $m=1$ warps \citep{widrow2019}, which are suggestive of the physics described in \cite{MT1997a, MT1997b}.
	
	The outline of this document is as follows. We begin in Section \ref{dynamicsBackground} with some background on collisionless self gravitating systems in a Hamiltonian framework to contextualize our proposed methodology. This is followed by a summary of Koopman theory, and the computational techniques employed in approximating the Koopman operator in Section \ref{datadriven}. We then provide an overview of the specific problem we will solve before proposing our solution in Section \ref{mapping}. Section \ref{examples} contains two applications of our method to test systems, first a simple quantum harmonic oscillator, and then the more astrophysically motivated isothermal plane model. Discussion of future work and concluding remarks are in Section \ref{conclusion}.  
	
	\section{Collisionless Boltzmann Equation} \label{dynamicsBackground}
	
	Consider a collisionless system with one spatial degree of freedom $ q $, conjugate momentum $ p $, and time parameter $ t $. For such a system, construct the continuous DF, $ f(q,p,t) $, and the corresponding configuration space density
	
	\begin{equation}\label{rho}
		\rho(q,t) = \int_{-\infty}^{+\infty}f(q,p,t)dp.
	\end{equation}
	
	\noindent The equations of motion in this system are derived form the single particle or mean field Hamiltonian  
	
	\begin{equation}\label{hamiltonian}
		H(q,p) = \tfrac{1}{2}p^2 + \Phi(q,t),
	\end{equation}
	
	\noindent where the potential $ \Phi(q,t) $ depends on the full configuration space distribution $ \rho(q,t) $ and external masses described by $ \Phi_\text{ext}(q) $. We take $ \rho(q,t) $ and $ f(q,p,t) $ to comprise a self-consistent solution to the collisionless Boltzmann Equation (CBE). That is, we have 
	
	\begin{equation}\label{cbe}
		\frac{\partial f}{\partial t} + \{f,H\} = \frac{\partial f}{\partial t} + p\frac{\partial f}{\partial q} - \frac{\partial \Phi}{\partial q}\frac{\partial f}{\partial p} = 0,
	\end{equation}
	
	\begin{equation}\label{poisson}
		\nabla^2\Phi(q,t) = 4\pi G\rho(q,t) + \nabla^2\Phi_{\text{ext}}(q),
	\end{equation}
	
	\noindent where $ G $ denotes the gravitational constant, and $ \{f,H\} $ the Poisson bracket of the functions $ f $ and $ H $. Since the potential depends on the DF, the system is nonlinear and there is no fully general prescription for generating solutions for either the time-dependent or time-independent cases. However, if an equilibrium solution is known, then one can  study time-dependent behavior analytically by
	considering the linearized CBE.
	
	
	For a known equilibrium DF, $ f_0(q,p) $, one may consider a perturbative solution
	
	\begin{equation}\label{perturbedDF}
		f(q,p,t) = f_0(q,p) + \epsilon f_1(q,p,t), \ \ |f_0|\simeq |f_1|\  \text{and} \ |\epsilon| \ll 1.
	\end{equation}
	
	\noindent Substituting this DF into Equation \ref{cbe} and dropping terms of order $ \epsilon^2 $ and higher, we obtain the linearized CBE, 
	
	\begin{equation}\label{lcbe}
		\frac{\partial f_1}{\partial t} + \{f_1,H_0\} + \{f_0,\Phi_1\} = 0.
	\end{equation}
	
	\noindent Solving equation \ref{lcbe} for $ f_1(q,p,t) $ yields a set of generally time-dependent modes of the DF. Additionally since equation \ref{cbe} is linear, given a set of modes $ f_j(q,p,t) $, each a solution of equation \ref{lcbe}, we can analyze any linear combination of the $ f_j(q,p,t) $ as a solution, and determine corresponding potential modes $ \Phi_j $ paired with the $ f_j $. We can then consider instead of the DF in equation \ref{perturbedDF}, one of the form
	
	\begin{equation}\label{lcbeDF}
		f(q,p,t) = f_0(q,p) +\epsilon\sum_{j>0} f_j(q,p,t).
	\end{equation}

	\section{Nonlinear Dynamics} \label{datadriven}
	
	When we describe a galaxy with $ f(q,p,t) $, we are using the abstraction of a phase space probability density to describe a more fundamental dynamical system, namely the evolution of all $ N $ stars as they interact with each other and external forces. A general form for the equations of motion of such a system is 
	
	\begin{equation}\label{statespace}
		\frac{d}{dt}\mathbf{x}(t) = \mathcal{T}(\mathbf{x}(t)),
	\end{equation} 
	
	\noindent where $ \mathbf{x} \in\mathbb{R}^{2N}$ is a \textit{state vector} containing the phase space coordinates of the $ N $ stars, and $ \mathcal{T}(x) $ is a generally nonlinear function determined by Hamilton's equations.  
	
	Since this system can be prohibitively complex for large numbers of stars, we opt for a statistical description in the form of probability density function, $ f(q,p,t) $, which determines the probability of finding a particle in a volume of phase space. The evolution of this function is defined by Hamilton's Equations and probability conservation. When we identify the evolution of $ \mathbf{x} $ with that of $ f(q,p,t) $, we are constructing a scalar observable function of the $ 2N $ dimensional phase space, that is $ f:\mathbf{x}\in\mathbb{R}^{2N}\mapsto\mathbb{R} $.

	\subsection{Koopman Theory}\label{koopman}
	
	Koopman theory facilitates producing a genuinely linear description of systems that are nonlinear in the space of their state vectors, here phase space. The system evolution is described in terms of observables \textit{on} the phase space rather than the phase space coordinates themselves. These observables evolve linearly in time according to the Koopman operator.
	
	At this point we transition to a discrete-time description both for notational convenience, and because we will ultimately work with discrete-time simulations. In terms of a discrete time variable $ t_m $, such that $ t_{m+1}=t_m+\Delta t $, the evolution of the state vectors is then
	
	\begin{equation}\label{nonlinear}
		\mathbf{x}(t_{m+1}) = T(\mathbf{x}(t_m)).
	\end{equation}
	
	\noindent Here, $ T $ is the discrete-time analog to $ \mathcal{T} $, and is in general non-linear. Suppose that we have an observable represented by a scalar function $ f:\mathbf{x}\mapsto\mathbb{R} $. This could be a density, or any other scalar function of $ \mathbf{x} $, but for our purposes the observable function is to be identified with the DF. The Koopman operator, $ \mathcal{K} $ maps $ f(\mathbf{x}) $ to another function $ \mathcal{K}f(\mathbf{x}) $ defined by the composition of  $ T(\mathbf{x}) $ and $ f(\mathbf{x}) $. Explicitly, we have the mapping
	
	\begin{equation}
		\mathcal{K}f(\mathbf{x}(t_m)) = f\circ T(\mathbf{x})= f(T(\mathbf{x}(t_m))).
	\end{equation}
	
	\noindent From equation \ref{nonlinear}, $ T(\mathbf{x}(t_m)) = \mathbf{x}(t_{m+1}) $, and therefore
	
	\begin{equation}\label{koopman_eq}
		\mathcal{K}f(\mathbf{x}(t_m)) = f(\mathbf{x}(t_{m+1})).
	\end{equation}
	
	\noindent That is, $ \mathcal{K}$  $ $ is a \textit{linear} evolution operator for the observable $ f(\mathbf{x}) $. Note that the Koopman operator acts on the space of all possible observables, of which $ f(\mathbf{x}) $ is a single element. The observable space is infinite dimensional and consequently so is the Koopman operator itself. So the nonlinear system in equation \ref{nonlinear} has become linear, with the caveat that the dynamics are represented in an infinite dimensional space. The advantage of describing the dynamics with equation \ref{koopman_eq} rather than equation \ref{nonlinear} lies in the eigenfunction expansion of linear operators. The evolution of the system can be represented in terms of the eigendecomposition, or spectrum, of the Koopman operator, and it is these eigenfunctions and eigenvalues we aim to estimate in our methodology. Note that we use eigendecomposition and spectrum interchangeably throughout this paper.
	
	\subsection{Dynamic Mode Decomposition}
	
	Since the exact Koopman operator is infinite dimensional, its explicit form cannot be determined in general. We instead compute finite dimensional approximations with an algorithm called Dynamic Mode Decomposition (DMD). This method determines the best fit linear evolution operator, along with its eigenfunctions and eigenvalues, for time series data. Here we provide a brief overview of DMD, for which a detailed introduction can be found in \cite{kutz2016}.    
	
	In what follows we assume a time series comprised of $ M $ snapshots of the scalar observable function $ f(q,p,t) $ arranged in the data matrix
	
	\begin{equation}\label{dataMatrix}
		\mathbf{F} = \begin{pmatrix}
			| & | &   & | \\
			\mathbf{f}(t_1) & \mathbf{f}(t_1) &  ... & \mathbf{f}(t_{M}) \\
			| & | &  &  | 
		\end{pmatrix},
	\end{equation}

	\noindent where each column, $ \mathbf{f}(t_m) $ is the DF sampled on a grid in $ (q,p) $ at time $ t_m $ and reshaped into a column vector. In DMD, one looks for series solutions of the continuous-time linear flow equation
	
	\begin{equation}\label{flowDMDcont}
		\frac{d}{dt}\mathbf{f}(t) = \mathcal{A}\mathbf{f}(t),
	\end{equation}
	
	\noindent that are the best fit for the discrete time data in $ \mathbf{F} $. Here the operator $ \mathcal{A} $ acts as a finite dimensional proxy for the Koopman operator. This operator possesses complex valued eigenfunctions, $ \psi_j $,  and eigenvalues, $ \omega_j $, that satisfy 
	
	\begin{equation}
		\mathcal{A}\psi_j=\omega_j\psi_j. 
	\end{equation}

	\noindent The general solution to equation \ref{flowDMDcont} can be written in terms of this eigendecomposition as

	\begin{equation}\label{seriesDF}
		f(q,p,t) =  \sum_{j=0}^{r-1} b_j\psi_j(q,p)\mathrm{e}^{\omega_jt},
	\end{equation}
	
	\noindent where the $ b_j $ are time-independent coefficients determined by the initial condition. The sum here runs over $ r $ terms, where $ r $ is the ``rank'' of the solution. We will discuss this more below. We remark that since $ \omega_j \in\mathbb{C} $, each term can contain oscillation, growth, and/or decay. The imaginary components of the frequencies can be seen as analgous to the typical Fourier frequencies, and there are circumstances in which there is a direct correspondence between a temporal discrete Fourier transform and DMD \citep{dft}.  
	
	To obtain these solutions in practice, one considers the corresponding discrete-time system
	
	\begin{equation}\label{flowDMD}
		\mathbf{f}(t_{m+1}) = \mathbf{A}\mathbf{f}(t_m),
	\end{equation}
	
	\noindent where the discrete-time and continuous-time operators are related by  
	
	\begin{equation}
		\mathbf{A} = \mathrm{e}^{\mathcal{A}\Delta t}.
	\end{equation}
	
	\noindent  The discrete-time operator has the eigendecomposition
	
	\begin{equation}\label{eigen}
		\mathbf{A}\psi_j=\lambda_j\psi_j,
	\end{equation}
	
	\noindent where 
	
	\begin{equation}\label{omega}
		\omega_j = \ln{(\lambda_j)}/\Delta t.
	\end{equation}
	
	For a time series of the form in equation \ref{dataMatrix}, the spectrum of the discrete-time operator is computed as follows. First construct two time-shifted matrices from $ \mathbf{F} $, defined as
	
	\begin{equation}\label{key}
		\mathbf{F}_1 = \begin{pmatrix}
			| &    & | \\
			\mathbf{f}(t_1) &  ... & \mathbf{f}(t_{M-1}) \\
			| &   &  | 
		\end{pmatrix},
		\mathbf{F}_2 = \begin{pmatrix}
			| &    & | \\
			\mathbf{f}(t_2) &  ... & \mathbf{f}(t_{M}) \\
			| &   &  | 
		\end{pmatrix}.
	\end{equation}
	
	\noindent The discrete-time evolution operator is then 
	
	\begin{equation}\label{key}
		\mathbf{A}=\mathbf{F}_2\mathbf{F}_1^+,
	\end{equation}
	
	\noindent where $ \mathbf{F}_1^+ $ denotes the Moore-Penrose pseudo inverse of $ \mathbf{F}_1 $. We calculate the pseudo inverse in terms of the singular value decomposition (SVD). For the SVD, $ \mathbf{F}_1\simeq \mathbf{U}\boldsymbol{\Sigma}\mathbf{V}^\dagger$, with $ ^\dagger $ the Hermitian conjugate transpose, $ \boldsymbol{\Sigma} $ a diagonal singular value matrix, and $ \mathbf{U} $ and $ \mathbf{V} $ left and right orthogonal matrices respectively. In terms of these quantities, the pseudo inverse is 
	
	\begin{equation}\label{key}
		\mathbf{F}_1^+=\mathbf{V}\boldsymbol{\Sigma}^{-1}\mathbf{U}^\dagger, 
	\end{equation}
	
	\noindent and the time evolution operator is 
	
	\begin{equation}\label{key}
		\mathbf{A}=\mathbf{F}_2\mathbf{V}\boldsymbol{\Sigma}^{-1}\mathbf{U}^\dagger.
	\end{equation}
	
	\noindent Instead of computing the eigendecomposition for $ \mathbf{A} $ directly, we work with its projection onto the space spanned by the $ r $ orthogonal vectors of $ \mathbf{U} $ corresponding to the $ r $ largest singular values in $ \boldsymbol{\Sigma} $. In what follows, all components of the SVD are truncated to rank $ r $. The projected time evolution operator is
	
	\begin{equation}\label{key}
		\tilde{\mathbf{A}}=\mathbf{U}^\dagger\mathbf{A}\mathbf{U}=\mathbf{U}^\dagger\mathbf{F}_2\mathbf{V}\boldsymbol{\Sigma}^{-1}.
	\end{equation}
	
	\noindent The rank is chosen at the outset of the procedure, and sets the dimension of the projected time evolution operator.  Choice of rank can be made rigorously by determining an optimal threshold on the singular values \citep{gavish2013}, or with DMD specific techniques, for example \cite{kou2017}. 
	
	
	We compute the eigenvectors and eigenvalues of $ \tilde{\mathbf{A}} $ by solving the equation
	
	\begin{equation}\label{key}
		\tilde{\mathbf{A}}\boldsymbol{\Xi}=\boldsymbol{\Xi}\boldsymbol{\Lambda}.
	\end{equation}
	
	\noindent The $ r $ eigenvalues of $ \tilde{\mathbf{A}} $ along the diagonal of $ \boldsymbol{\Lambda} $ are eigenvalues of $ \mathbf{A} $. These eigenvalues are ordered according to their corresponding singular values in the SVD. The eigenvectors of $ \mathbf{A} $ however are obtained from \citep{kutz2016} 
	
	\begin{equation}\label{key}
		\boldsymbol{\Psi}=\begin{pmatrix}
			| & &  | \\
			\psi_0 &...  & \psi_{r-1} \\
			| & & |  \\
		\end{pmatrix}=\mathbf{F}_2\mathbf{V}\boldsymbol{\Sigma}^{-1}\boldsymbol{\Xi}.
	\end{equation}
	
	\noindent The time-dependent coefficients, $ b_j $, are then computed from the initial condition as

	\begin{equation}\label{key}
		\mathbf{b}=(b_0\ ...\ b_{r-1})^T=\boldsymbol{\Psi}^+\mathbf{f}(t_0),
	\end{equation}	
	
	\noindent where $\boldsymbol{\Psi}^+ $ denotes the Moore-Penrose pseudo-inverse of $ \boldsymbol{\Psi} $.

	In summary, when we apply DMD with state vectors mapped into observables we are treating the linear evolution operator $ \mathbf{A} $ as a finite-dimensional, rank-$ r $ approximation of the Koopman operator. Further, the spectrum of  $ \mathbf{A} $ is known to be related to that of $ \mathcal{K} $ \citep{Tu2013}, and under particular conditions, the two operators share eigenvalues \citep{kutz2016}.
	
	\subsection{Koopman and Linear Perturbation Theory}
	
	As is the case in linear perturbation theory, the DF's dependence on time and phase space is separated in equation \ref{seriesDF}. This solution is however fundamentally different than equation \ref{lcbeDF}.  To apply perturbation theory to collisionless dynamics, one must first determine an equilibrium solution to the CBE. This solution generates a linear operator, which replaces the original nonlinear operator in equation \ref{cbe}. The resulting linear system is an approximation that is only valid when perturbations are small relative to the equilibrium solution. Conversely, when we ``linearize'' a system with the Koopman approach, we are not constructing a linearized operator dependent on a zeroth order solution, but rather transforming the system to a higher dimensional space which naturally admits a linear representation of the system. This leads to two important features of equation \ref{seriesDF}. First, that there is no condition on the magnitude of the time-dependent terms relative to the equilibrium model (the $\omega_j = 0$ terms). Second, that the equilibrium model and the time-dependent terms are determined simultaneously, not requiring prior knowledge of a time-independent solution to equation \ref{cbe}.
	
	\section{Inferring Time-Dependence}\label{mapping}
	
	In this Section, we propose a method for inferring a time-dependent model from a single phase space snapshot. This inference problem can be stated as estimating the function $ G:f(q,p;t_0)\mapsto f(q,p,t) $ that maps a snapshot at time $ t_0 $ to the time-dependent DF. The semicolon here indicates that the fixed time, $ t_0 $, is a parameter of the snapshot, which is itself not a function of time. The goal is to construct an appropriate non-parametric model for the function $ G $. Such a model is trained on a set of simulated systems to facilitate making inferences from snapshots sampled from an unknown DF. We will use CNNs to approximate the mapping $ G $.
	
	\subsection{Mapping Snapshots to the Koopman Spectrum}
	
	The DMD series solution plays a crucial role of ``compression” in our methodology. When determining a rank-$ r $ DMD solution for a sequence of $ M $ snapshots, each with grid resolution $ M_q\times M_p $ (where we always have $ r\ll M $), we are storing all of the information that the rank-$ r $ decomposition is able to capture into $ r $ eigenfunctions, and $ r $ eigenvalues. This is $ r(1+M_qM_p) $ complex numbers, in contrast to the $ MM_qM_p $ real numbers needed to store the entire simulation. Additionally, the eigenvalues and eigenfunctions  come in conjugate pairs, so we need only store half of them. Note that the coefficients in the DMD series can be obtained by projecting $ f(q,p;t_0) $ onto the eigenfunctions. Therefore, they do not need to be stored.
	
	Since the spatiotemporal behavior of the DF is captured in the $ r $ eigenfunctions and eigenvalues, we can map the snapshot $ f(q,p;t_0) $ to these quantities instead of an explicit time-series. Additionally, it will be easier to design two separate models to perform mappings for the eigenvalues and eigenvectors separately, instead of having one model that attempts to map $ f(q,p;t_0) $ to the full spectrum. The input of the model is a snapshot of the DF, $ f(q,p;t_0) $, which acts as a Koopman observable of the state space, and the output is the eigendecomposition of a finite dimensional approximation of the Koopman operator. From this point on we will separate the problem into two functions, $ G_\omega:f(q,p;t_0)\mapsto \boldsymbol{\Omega}$ and $ G_\psi:f(q,p;t_0)\mapsto\boldsymbol{\Psi} $. That is, $ G_\psi $ maps the snapshot to the static phase space structures that form a basis for our predicted DFs, and $ G_\omega $ determines the temporal characteristics in the form of damping coefficients and frequencies. In Section \ref{examples} we will use separate neural network models for each of these mappings. 
	
	By splitting the mapping into two separate problems, we effectively break the relationship between $ \psi $ and $ \omega $. That is, in general, these quantities no longer obey equation \ref{eigen} and should perhaps not be called eigenvectors and eigenvalues. Regardless, $ \psi $ and $ \omega $ are, respectively, basis functions and complex-valued frequencies from which a time-dependent DF may be recovered.  Since our model is designed to estimate these quantities as eigenvectors and eigenvalues via DMD, we continue to use these terms. 
	
	In principle, one can avoid this problem by training a model to map kinematic snapshots to the operator $ \mathbf{A} $, which serves as a finite dimensional approximation of the Koopman operator. Genuine eigenvectors and eigenvalues could then be computed directly. In practice though, $ \mathbf{A} $ is a large $ (M-1)\times (M-1) $ matrix, of which the particular size depends on the number of time steps in a simulation. So, we lose the previously described compression, and have a model output that depends on parameters of the simulations used to make data, specifically the number of snapshots or temporal resolution. Furthermore, the eigendecomposition can become expensive in this approach (it is not computed directly in DMD). Network design is also made difficult by the variable output space dimension. Our contention is that $ \psi $ and $ \omega $ will do a better job of capturing the underlying dynamics of a system than the full evolution operator, which will depend on parameters of the simulation used to make the data. In either case, the efficacy of the model is contingent on the quality of the training set, and our splitting of the mappings to eigenvectors and eigenvalues may exacerbate this.
	
	The rank, $ r $, which controls the complexity of the model, must be chosen when designing the network and constructing training data. In the work presented here, $ r $ is a fixed parameter of $ G $. Were we to allow $ r $ to vary, we would not be able to use the simple non-parametric techniques described here. In the present work we are most interested in dominant, persisting structure rather than the weak persisting contributions and transients that tend to arise in higher rank DMD models. We therefore choose to work with a low rank. In the case that transients are to be included however, multi-resolution DMD \citep{mrDMD} and higher rank models can facilitate incorporating them into this framework. For an example of how relatively few terms dominate the sum in a nonlinear astrophysical system, we refer the reader to \cite{darling2019}. There one can see a comparison of DMD solutions to simulation data for the isothermal plane and homogeneous slab models. Mode dominance is visualized in the form of time-integrated amplitudes and mode lifetimes.
	
	\subsection{Convolutional Neural Networks}\label{CNNsection}
	
	Some basic knowledge of convolutional networks is assumed in Section \ref{examples}, however the essential background on this topic is covered here, with supplemental information in Section \ref{CNNbackground}. For a comprehensive review of CNNs, we refer the reader to \cite{goodfellow2016}.
	
	We realize the function $ G $ as a non-parametric regression model implemented with  CNNs. Koopman theory was previously combined with neural networks by \cite{lusch2018} who used them to determine the mapping from system coordinates to Koopman observables. To our knowledge, neural networks have never before been used to map single snapshots from time series to eigenfunctions and eigenvalues from which time evolution of a system can be constructed. 
	
\begin{figure}
		\resizebox{0.525\textwidth}{!}{%
			\begin{tikzpicture}[x={(-0.7071067811865475cm,-0.7071067811865475cm)},
				y={(1cm,0cm)}, z={(0cm,1cm)}]
				\coordinate (O) at (0, 0, 0);
				
				\begin{scope}[xshift=0,yshift=0]
					\draw[black]
					(0,0.1,0)--(0,0.1,3)--(0,0,3)--(0,0,0)--(0,0.1,0)
					(0,0,3)--(-3,0,3)--(-3,0.1,3)--(0,0.1,3)
					(0,0.1,0)--(-3,0.1,0)--(-3,0.1,3);
					\node [black,xshift=38,yshift=74]{\footnotesize $x$};
					\node [black,xshift=-30,yshift=60]{Input };
					\node [black,xshift=-30,yshift=50]{\footnotesize Image};
					\node [black,xshift=-30,yshift=40]{\footnotesize $(60\times60\times1)$};
				\end{scope}

				\begin{scope}[xshift=70,yshift=130]
					\draw[black]
					(0,0.1,0)--(0,0.1,0.5)--(0,0,0.5)--(0,0,0)--(0,0.1,0)
					(0,0,0.5)--(-0.5,0,0.5)--(-0.5,0.1,0.5)--(0,0.1,0.5)
					(0,0.1,0)--(-0.5,0.1,0)--(-0.5,0.1,0.5);
					\draw[black, thick, ->](0,0.3,0.4) to[out=0,in=180] (0,0.7,0.4);
					\node [black,xshift=35,yshift=12]{\footnotesize $x*k_1$};
					\draw[black](0,0.01,0.3)--(0,-1,-1.9);
					\node [black,xshift=6,yshift=-5]{\footnotesize $k_1$};
				\end{scope}
				
				\begin{scope}[xshift=70,yshift=90]
					\draw[black]
					(0,0.1,0)--(0,0.1,0.5)--(0,0,0.5)--(0,0,0)--(0,0.1,0)
					(0,0,0.5)--(-0.5,0,0.5)--(-0.5,0.1,0.5)--(0,0.1,0.5)
					(0,0.1,0)--(-0.5,0.1,0)--(-0.5,0.1,0.5);
					\draw[black, thick, ->](0,0.3,0.4) to[out=0,in=180] (0,0.7,0.4);
					\node [black,xshift=35,yshift=12]{\footnotesize $x*k_2$};
					\draw[black](0,0.01,0.3)--(0,-1,-0.5);
					\node [black,xshift=6,yshift=-5]{\footnotesize $k_2$};
				\end{scope}
				
				\begin{scope}[xshift=70,yshift=50]
					\draw[black]
					(0,0.1,0)--(0,0.1,0.5)--(0,0,0.5)--(0,0,0)--(0,0.1,0)
					(0,0,0.5)--(-0.5,0,0.5)--(-0.5,0.1,0.5)--(0,0.1,0.5)
					(0,0.1,0)--(-0.5,0.1,0)--(-0.5,0.1,0.5);
					\draw[black, thick, ->](0,0.3,0.4) to[out=0,in=180] (0,0.7,0.4);
					\node [black,xshift=35,yshift=12]{\footnotesize $x*k_3$};
					\draw[black](0,0.01,0.3)--(0,-1,0.9);
					\node [black,xshift=6,yshift=-5]{\footnotesize $k_3$};
				\end{scope}
				
				\begin{scope}[xshift=70,yshift=10]
					\draw[black]
					(0,0.1,0)--(0,0.1,0.5)--(0,0,0.5)--(0,0,0)--(0,0.1,0)
					(0,0,0.5)--(-0.5,0,0.5)--(-0.5,0.1,0.5)--(0,0.1,0.5)
					(0,0.1,0)--(-0.5,0.1,0)--(-0.5,0.1,0.5);
					\draw[black, thick, ->](0,0.3,0.4) to[out=0,in=180] (0,0.7,0.4);
					\node [black,xshift=35,yshift=12]{\footnotesize $x*k_4$};
					\draw[black](0,0.01,0.3)--(0,-1,2.3);
					\node [black,xshift=6,yshift=-5]{\footnotesize $k_4$};
					\node [black,xshift=5,yshift=-15]{\footnotesize Filter};
					\node [black,xshift=5,yshift=-25]{\footnotesize Kernels};
					\node [black,xshift=5,yshift=-35]{\footnotesize $ (3\times3\times1) $};
				\end{scope}

				\begin{scope}[xshift=120,yshift=0]
					\draw[black]
					(0,0.1,0)--(0,0.1,1)--(0,0,1)--(0,0,0)--(0,0.1,0)
					(0,0,1)--(-1,0,1)--(-1,0.1,1)--(0,0.1,1)
					(0,0.1,0)--(-1,0.1,0)--(-1,0.1,1);
					\draw[black, thick, ->](0,0.5,0.8) to[out=0,in=180] (0,1.,0.8);
					\node [black,xshift=55,yshift=23]{\footnotesize $\rm{ReLU}(x*k_4)$};
					\node [black,xshift=5,yshift=-5]{\footnotesize Convolution};
					\node [black,xshift=5,yshift=-15]{\footnotesize Output};
					\node [black,xshift=5,yshift=-25]{\footnotesize $ (58\times58\times1) $};
				\end{scope}

				\begin{scope}[xshift=120,yshift=40]
					\draw[black]
					(0,0.1,0)--(0,0.1,1)--(0,0,1)--(0,0,0)--(0,0.1,0)
					(0,0,1)--(-1,0,1)--(-1,0.1,1)--(0,0.1,1)
					(0,0.1,0)--(-1,0.1,0)--(-1,0.1,1);
					\draw[black, thick, ->](0,0.5,0.8) to[out=0,in=180] (0,1.,0.8);
					\node [black,xshift=55,yshift=23]{\footnotesize $\rm{ReLU}(x*k_3)$};
				\end{scope}
				
				\begin{scope}[xshift=120,yshift=80]
					\draw[black]
					(0,0.1,0)--(0,0.1,1)--(0,0,1)--(0,0,0)--(0,0.1,0)
					(0,0,1)--(-1,0,1)--(-1,0.1,1)--(0,0.1,1)
					(0,0.1,0)--(-1,0.1,0)--(-1,0.1,1);
					\draw[black, thick, ->](0,0.5,0.8) to[out=0,in=180] (0,1.,0.8);
					\node [black,xshift=55,yshift=23]{\footnotesize $\rm{ReLU}(x*k_2)$};
				\end{scope}
				
				\begin{scope}[xshift=120,yshift=120]
					\draw[black]
					(0,0.1,0)--(0,0.1,1)--(0,0,1)--(0,0,0)--(0,0.1,0)
					(0,0,1)--(-1,0,1)--(-1,0.1,1)--(0,0.1,1)
					(0,0.1,0)--(-1,0.1,0)--(-1,0.1,1);
					\draw[black, thick, ->](0,0.5,0.8) to[out=0,in=180] (0,1.,0.8);
					\node [black,xshift=55,yshift=23]{\footnotesize $\rm{ReLU}(x*k_1)$};
				\end{scope}

				\begin{scope}[xshift=205,yshift=0]
					\draw[black]
					(0,0.1,0)--(0,0.1,1)--(0,0,1)--(0,0,0)--(0,0.1,0)
					(0,0,1)--(-1,0,1)--(-1,0.1,1)--(0,0.1,1)
					(0,0.1,0)--(-1,0.1,0)--(-1,0.1,1);
					\draw[black](0,0.5,0.8)--(0,2,2.8);
					\node [black,xshift=5,yshift=-5]{\footnotesize Activation Function};
					\node [black,xshift=5,yshift=-15]{\footnotesize Output};
					\node [black,xshift=5,yshift=-25]{\footnotesize $ (58\times58\times1) $};
				\end{scope}

				\begin{scope}[xshift=205,yshift=40]
					\draw[black]
					(0,0.1,0)--(0,0.1,1)--(0,0,1)--(0,0,0)--(0,0.1,0)
					(0,0,1)--(-1,0,1)--(-1,0.1,1)--(0,0.1,1)
					(0,0.1,0)--(-1,0.1,0)--(-1,0.1,1);
					\draw[black](0,0.5,0.8)--(0,2,1.4);
				\end{scope}
				
				\begin{scope}[xshift=205,yshift=80]
					\draw[black]
					(0,0.1,0)--(0,0.1,1)--(0,0,1)--(0,0,0)--(0,0.1,0)
					(0,0,1)--(-1,0,1)--(-1,0.1,1)--(0,0.1,1)
					(0,0.1,0)--(-1,0.1,0)--(-1,0.1,1);
					\draw[black](0,0.5,0.8)--(0,2,0.);
				\end{scope}
				
				\begin{scope}[xshift=205,yshift=120]
					\draw[black]
					(0,0.1,0)--(0,0.1,1)--(0,0,1)--(0,0,0)--(0,0.1,0)
					(0,0,1)--(-1,0,1)--(-1,0.1,1)--(0,0.1,1)
					(0,0.1,0)--(-1,0.1,0)--(-1,0.1,1);
					\draw[black](0,0.5,0.8)--(0,2,-1.4);
				\end{scope}

				\begin{scope}[xshift=262,yshift=60]
					\draw[black]
					(0,0.4,0)--(0,0.4,1)--(0,0,1)--(0,0,0)--(0,0.4,0)
					(0,0,1)--(-1,0,1)--(-1,0.4,1)--(0,0.4,1)
					(0,0.4,0)--(-1,0.4,0)--(-1,0.4,1);
					\node [black,xshift=25,yshift=-5]{\footnotesize Convolutional Block};
					\node [black,xshift=25,yshift=-15]{\footnotesize Output};
					\node [black,xshift=25,yshift=-25]{\footnotesize $ (58\times58\times4) $};
				\end{scope}
				
			\end{tikzpicture}
		}%
		\caption{Example of a single convolutional block for a $ 60\times 60 $ input matrix (this could be a snapshot of a $ 2 $-dimensional DF on $ (q,p) $ sampled on a $ 60\times 60 $ grid.). The block consists of four $ 3\times 3  $ filters, and a rectified linear unit (ReLU) activation function, as defined in equation  \ref{twolayer2}.}
		\label{convblock}
	\end{figure}
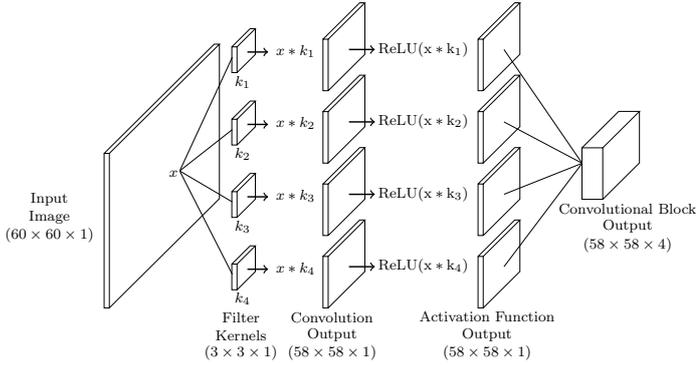

	For an $ N $-dimensional input space the learnable parameters of the CNN are the elements of $ N $-dimensional filters that are used in convolution operations. The CNN comprises a series of layers. The mapping performed in each layer is the convolution of the input of the layer with that layer's filters, followed by the application of an activation function to the result of the convolution. A simple concrete example of this for a $ 2 $-dimensional input is visualized in Fig. \ref{convblock}. This diagram demonstrates the signal path of the input matrix through to the output, and the exact role the learnable filters play in the mapping performed by the block. We cascade multiple blocks of this form in the models we use in Section \ref{examples}, and combine them with down-sampling and reshaping techniques to obtain our desired output space dimension. 
	
	For our purposes, CNNs have two main advantages over the more conventional multi-layer perceptrons (MLPs, see \ref{MLPbackground}), which use linear transformation matrices instead of convolution. First, CNNs share parameters among multiple operations. This makes training and evaluating models involving large data volumes more efficient than for the equivalent MLP. The examples in Section \ref{examples} are of low dimension and could be handled by MLPs. However, it would be prohibitively expensive to scale up the resulting models for large spaces like the $ 6 $-dimensional phase space required for complete modeling of a galaxy. CNNs provide a comparatively feasible route forward in scaling to higher dimension. Second, CNNs have a natural affinity for identifying spatial structure in data, which comes from their use of convolution. Consider for example a function that maps an input vector $ \mathbf{x}$ into an output vector $ \mathbf{y} $ via either a linear transformation (matrix), $ \mathbf{W} $, or convolution with a filter kernel $ \mathbf{k} $. In the case of convolution, the output $ \mathbf{y} $ is manifestly dependent on the \textit{arrangement} of the elements in $ \mathbf{x} $ at the scale of the number of elements in the kernel. Conversely for the linear transformation, without very particular constraints on the structure of $ \mathbf{W} $ (Toeplitz form, for example), the output does not depend on local structure in $ \mathbf{x} $.
	
	All neural network models used in section \ref{examples} are implemented with basic TensorFlow \citep{tensorflow2015} tools. The network architectures, and all parameters are summarized in the appendix, in tables \ref{TableSHO}, \ref{vectorTable}, and \ref{valueTable}.
	
	\section{Example Models}\label{examples}
	
	Here we apply the methodology described in the previous sections to two toy model problems. First, in Section \ref{qsho_section},  we consider the Schr\"{o}dinger equation with a harmonic potential. This is followed by the isothermal plane model (or Spitzer sheet) in Section \ref{isothermal_section}, where we must use the methods described in Section \ref{datadriven} to search for transformations to a space in which the dynamics become linear.

	\subsection{Quantum Simple Harmonic Oscillator}\label{qsho_section}
	
	We begin with the quantum harmonic oscillator for two reasons. First, as the Schr\"{o}dinger equation is linear, its solutions may always be written exactly as eigenfunction expansions of the form in equation \ref{seriesDF}. Additionally, since the Hamiltonian is a Hermitian operator, the eigenvalues are strictly real, which simplifies the regression problem involved in determining the mapping from snapshot to eigenvalues. In the particular case of a harmonic potential, we have analytic expressions for the eigenfunctions and eigenvalues, so we can omit the DMD step of our procedure and focus solely on the problem of inferring system dynamics from a single snapshot with a non-parametric model. Further, the Schr\"{o}dinger equation admits a probabilistic description of configuration space dynamics, converse to the full phase space needed in the case of the CBE. This means the input and output spaces are significantly smaller than in the collisionless dynamics case. 
	
	This example is not entirely unrelated to stellar dynamics. In fact, the Schr\"{o}dinger equation has been applied to collisionless self-gravitating systems in \cite{widrowKaiser1993}. In that paper, the authors constructed a system comprised of the Schr\"{o}dinger and Poisson equations such that the typical coupled collisionless Boltzmann and Poisson equations are recovered in a particular limit. 
	
	\subsubsection{System and Data}
	
	Consider observations of a wavefunction that satisfies the Schr\"{o}dinger equation with a harmonic potential.  For convenience we take $\hbar=m=1 $, and write
	
	\begin{equation}\label{schrodinger}
		\frac{\partial}{\partial t}\Psi(q,t) = -\mathrm{i}\mathcal{H}\Psi(q,t), \ \ \mathcal{H}=-\frac{1}{2}\frac{\partial^2}{\partial q^2} + \frac{\xi}{2}q^2.
	\end{equation}
	
	\noindent Suppose that the potential curvature $ \xi $ is unknown. The problem then becomes: given one time series sample $ \Psi(q;t_0) $ from the evolution of a wavefunction, with unknown initial condition, and potential curvature $ \xi $, can we determine past and future states of the wavefunction. Although we do not know the curvature $ \xi $, we will assume that it belongs to a normal distribution of possible values with mean, $ \bar{\xi} $ and variance $ \sigma_\xi^2 $,  denoted $ N(\bar{\xi},\sigma_\xi) $. The idea here is that in order for one to efficiently produce training data, and subsequently train a model for use on a system, some knowledge of system parameters must be known. 
	
	As equation \ref{schrodinger} is a linear time evolution equation, we have the familiar general solution obtainable through separation of variables, 
	
	\begin{equation}\label{wavefunction}
		\Psi(q,t) = \sum_{j=0}^{\infty}a_j\psi_j(q)\mathrm{e}^{-\mathrm{i}E_jt},
	\end{equation} 
	
	\noindent where the eigenfunctions $ \psi_j(q) $ and eigenvalues $ E_j $ satisfy the linear differential eigenvalue equation
	
	\begin{equation}\label{tidse}
		\mathcal{H}\psi_j(q)=E_j\psi_j(q).
	\end{equation}
	
	\noindent To be consistent with Section \ref{datadriven},  we define the complex frequencies 
	
	\begin{equation}\label{key}
		\omega_j=-\mathrm{i}E_j. 
	\end{equation}

	\noindent The eigenfunctions of equation \ref{schrodinger} are well known to be 
	
	\begin{equation}\label{sho_eigenfunctions}
		\psi_j(q) = \bigg(\frac{\xi}{\pi}\bigg)^{1/4}\frac{1}{\sqrt{2^jj!}}H_j\bigg(\sqrt{\xi q^2}\bigg)\mathrm{e}^{-\frac{\xi}{2}q^2},
	\end{equation}
	
	\noindent where $ H_j(q) $ are the Hermite polynomials of order $ j $. The corresponding eigenvalues are given by
	
	\begin{equation}\label{sho_eigenvalues}
		E_j = \big(j+\tfrac{1}{2}\big) \xi.
	\end{equation}
	
	The input-output pairs of the mapping $ G $ are, $ (\Psi(q;t_0),(\psi_j,\omega_j)) $, where the input, $ \Psi(q;t_0) $, is an evolved state of a random initial condition. The output, $ (\psi_j,\omega_j) $, is the eigendecomposition of the Hamiltonian operator. For each element of the training data set, we (i) sample $ \xi $ from $ N(\bar{\xi},\sigma_\xi) $; (ii) generate a random initial condition comprised of the ground state in superposition with excited states assigned random coefficients; (iii) determine the eigenvectors and eigenvalues, $ (\psi_j,\omega_j) $, for the chosen $ \xi $; and (iv) evolve the initial condition to a randomly selected time $ t_0 $, to produce $ \Psi(q;t_0)  $. For the purposes of our calculations, we approximate the eigenfunctions by evaluating equation \ref{sho_eigenfunctions} on a $ 500 $ point grid in $ q $. This means that the non-parametric model sees the outputs in the training set simply as vectors, and has no prior knowledge of the explicit form of the eigenfunctions, namely that they contain Hermite polynomials. 
	
	For the training data we sample $ 5\times 10^4 $ potentials from $ N(\bar{\xi},\sigma_\xi) $ with mean potential curvature $ \bar{\xi}=5 $, and variance $ \sigma_\xi^2=1 $. We use a test set of $ 10^5 $ points, with values of $ \xi $ sampled from a uniform distribution, $ U(\bar{\xi}-3\sigma_\xi, \bar{\xi}+3\sigma_\xi) $. This means the model is tested on approximately the same number of points for each value of $ \xi $ within the range of the uniform distribution facilitating investigation into how well the model generalizes to systems away from $ \xi=\bar{\xi} $. This will be made clear when we evaluate the model performance in Figs. \ref{e_dist_vec} and \ref{e_dist_val}. In both cases, the input-output pairs are generated from the sampled potentials as described in the previous paragraph.

	\subsubsection{Neural Network Model and Results}
	
	For this problem we use simple one dimensional CNNs. Following the discussion in Section \ref{mapping}, we split the problem of determining the mapping $ G $ into the two sub-problems described by $ G_\psi:\Psi(q;t_0)\mapsto \psi_j $ and $ G_\omega:\Psi(q;t_0)\mapsto \omega_j $, each implemented by their own CNN. Both models use the same convolutional structure, but differ in the sizes of their outputs and hyperparameters. The shared network architecture is shown in Table \ref{TableSHO}.
	
	The eigenvector predictor, $ G_\psi $ is trained with a learning rate of $ 1\times 10^{-5} $, and a patience parameter of $ 50 $. Evaluated on the test data, the trained model has a mean squared error of $  4.7\times10^{-5} $, which corresponds to $\sim 0.05\% $ of the mean eigenvector magnitude. The eigenvalue predictor is trained with a learning rate of $ 1\times 10^{-4} $ and patience of $  100$. This yields a mean squared error of $ 0.018 $ on the test data, which is $ \sim0.2\% $ of the average eigenvalue magnitude.

	\begin{figure}
		\centering
		\includegraphics[width=9.9cm]{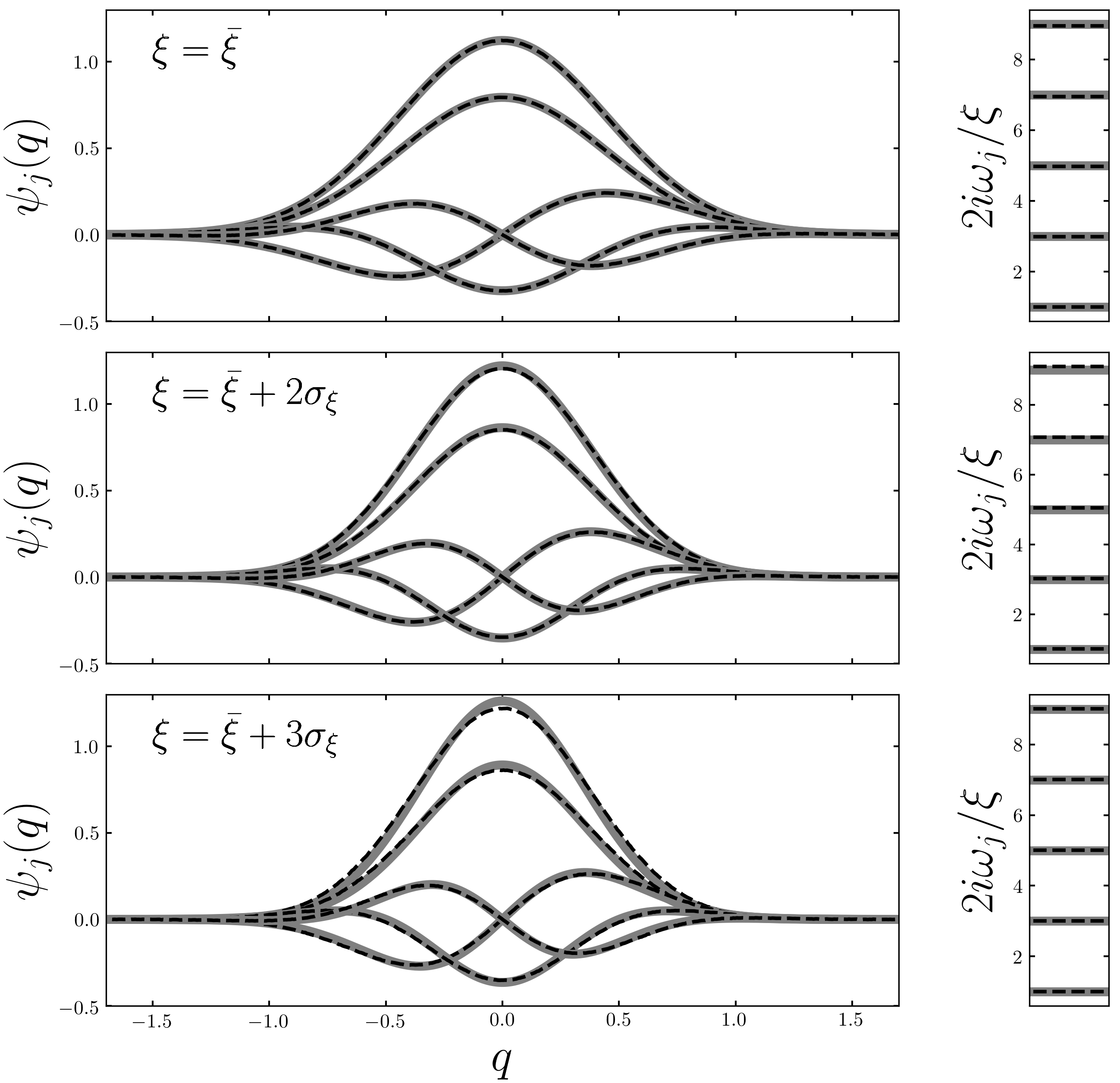}
		\caption{Comparison of true eigenvectors and eigenvalues to model predictions. The eigenvectors are shown in the large left-side panels, and the eigenvalues in the small right-side panels. True values and model predictions are shown as gray solid lines and black dotted lines respectfully.}
		\label{SHOresults}
	\end{figure}
	
	Fig. \ref{SHOresults} shows representative examples of model performance evaluated on the test data. The values of $ \xi $ shown in the three panels here correspond to the mean value of the parameter, along with values exactly two and three standard deviations away from the mean.
	
\begin{figure}
		\includegraphics[width=8.99cm]{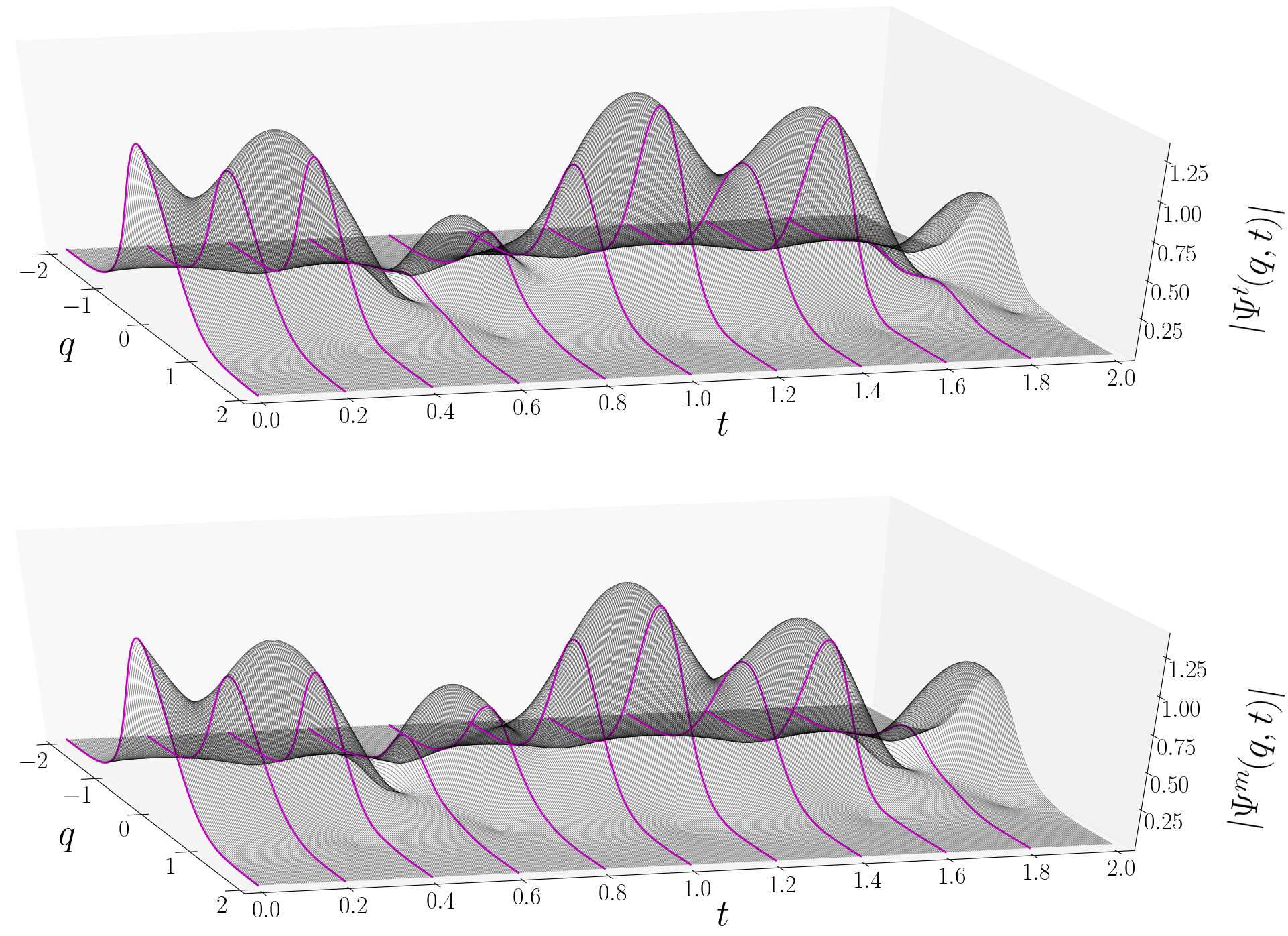}
		\caption{Time evolution of the true (top) and model output (bottom) wave functions. Each thin black curve is a time series sample from their evolution. The thick magenta curves are drawn at much larger sampling intervals to emphasize how the two wave functions differ at identical times.}
		\label{SHOresults2}
	\end{figure}

	The models are successful in estimating both the eigenfunctions and eigenvalues. Note that although the eigenvalues are predicted with relatively high accuracy, we should not expect them to exactly match the true values. Since the eigenvalues are arguments of temporal exponential functions, small errors lead to large changes in the wavefunction as time progresses. That is, although the wave function has the correct qualitative behavior, it goes out of phase with the exact solution on a time scale proportional to the inverse of the error in the eigenvalue. This point is illustrated in Fig. \ref{SHOresults2} where we show the time-dependence of the true and predicted wave functions for the test case from the first row of Fig. \ref{SHOresults}. The predicted wave function is in good qualitative agreement with the true one though it is clearly out of phase by $ t\simeq1.8 $ as indicated by the right-most magenta curve.

	\begin{figure}
		\centering
		\includegraphics[width=8cm]{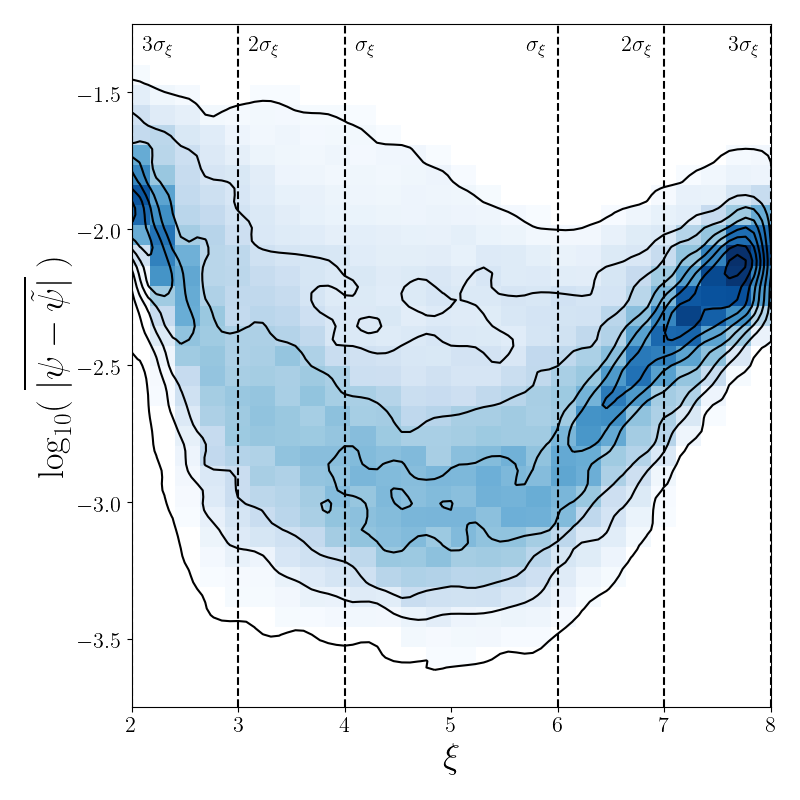}
		\caption{Density of test points in the space of the model parameter and mean absolute prediction error on the eigenvectors. The blue histogram is normalized density in this space, and the black curves are corresponding contours. Dashed lines indicate how far a test point's corresponding $ \xi $ value is from the mean in one $ \sigma_\xi $ increments.}\label{e_dist_vec}
	\end{figure}

	\begin{figure}
		\centering
		\includegraphics[width=8cm]{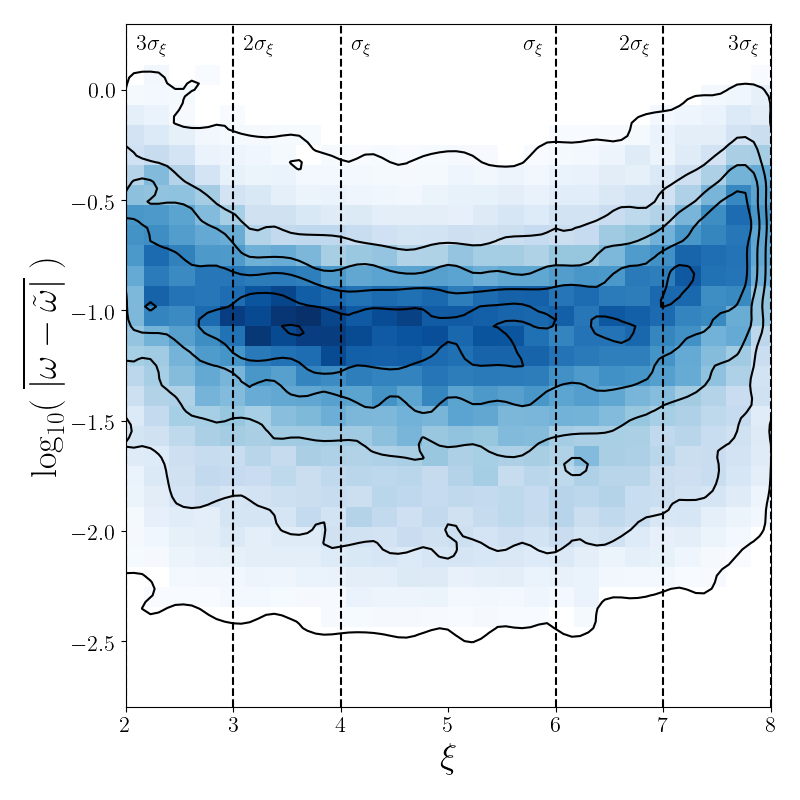}
		\caption{Same as Fig. \ref{e_dist_vec} but for the eigenvalue predictor.}\label{e_dist_val}
	\end{figure}
	
	Fig. \ref{e_dist_vec} and \ref{e_dist_val} indicate how well the models generalize away from the training data. These figures show the density of points from the test set in the space of $ \xi $ and average absolute error on predictions for $ \psi $ and $ \omega $. Here, the average absolute error is defined $ \overline{|\psi-\tilde{\psi}|}=\frac{1}{M}\sum_{i=1} |\psi(x_i)-\tilde{\psi}(x_i)|$. For both models, the error increases as $ \xi $ values approach the tails of the distribution $ N(\bar{\xi},\sigma_\xi) $, as is consistent with the expectation from exposing the model to more points close to $ \bar{\xi} $ than away from it.

	Note that the error distributions for $ G_\psi $ and $ G_\omega $ differ greatly. The eigenvalue model, shown in Fig. \ref{e_dist_val}, has what looks to be a single elongated region of high density, which aligns with a roughly quadratic dependence on $ \xi $. Additionally, the distribution in error at any fixed value of $ \xi $ has approximately the same spread as at any other $ \xi $ value, as can be seen in the contours. Conversely, the eigenvector distribution shown in Fig. \ref{e_dist_vec}, has two local maxima at the edges of the $ \xi $ range. These result from the spread in error being much tighter at $ \xi $ further than $ 2\sigma_\xi $ away from $ \bar{\xi} $, than $ \xi $ near $ \bar{\xi} $. The takeaway from this is that the upper and lower bounds on the error in the eigenvalue predictor are relatively independent of $ \xi $, compared to the case of the eigenvector predictor, where the error bounds differ greatly near $ \bar{\xi} $, and converge as one moves towards the tails of $ N(\bar{\xi},\sigma_\xi) $. This indicates that the eigenvector predictor is more susceptible to error in the regions on the parameter space it has not been heavily trained on than the eigenvalue predictor. It is possible that this is simply a result of the more complex nature of the eigenvector problem, $ G_\psi:\Psi(q;t_0)\mapsto \psi_j $, with its larger output space compared to the eigenvalue problem, $ G_\omega:\Psi(q;t_0)\mapsto \omega_j $. Whether this can be effectively remedied simply with further training as opposed to modifications to network architecture is currently unclear and not within the scope of this work. It should be noted however that in the framework as it is presented, eigenvector predictors are not as robust to incorrect assumptions on system parameters as their eigenvalue counterparts. 
	
	In Fig.  \ref{e_dist_vec}, the error in $ \psi $ exhibits asymmetry in $\xi $, with errors rising faster for $ \xi<\bar{\xi} $. A possible explanation for this is that the dynamical time of the system depends on $ \xi $, with $ \xi>\bar{\xi} $ realizations having shorter orbital periods than their $ \xi<\bar{\xi} $ counterparts. This results in the model seeing effectively less information about the dynamics for $ \xi<\bar{\xi} $ in the case of unbiased sampling from $ N(\bar{\xi},\sigma_\xi) $. This is because information of the potential is encoded in the evolved state as a result of the wave function and potential interacting over time.
	
	We end this section with a remark that with the knowledge required to produce this training data, one could potentially learn the spectrum of the system from a single snapshot by simply assuming a dominant ground state and fitting a Gaussian for the $ j=0 $ mode. The curvature of the potential could then be inferred without the need for this framework. The purpose of this example is not to revolutionize the way we solve for time-dependence of the harmonic oscillator, but rather to take a simple system, and see if a non-parametric model is capable of accurately mapping snapshot to spectrum. A more interesting problem would be to consider the Schr\"{o}dinger equation with an unknown functional form of the potential, where we train the model on simulations for many different potentials and initial conditions. Since this problem approaches the complexity of the self-gravitating systems we aim to study, we move on to a system described by the collisionless Boltzmann and Poisson equations. 
	
	\subsection{Modified Isothermal Plane}\label{isothermal_section}
	
	The isothermal plane model was developed by \cite{spitzer1942} and \cite{camm1950} and used to study the vertical structure of a localized patch in a stellar disc in \cite{freeman1978} and \cite{vanderkruit1981}. \cite{darling2019} introduced a simple modification of this model in which a fraction of the gravitational potential is assumed to be static, while the remainder comes from the live particles in the patch. In essence, the static potential is meant to account for
	distant parts of the disk, the dark halo, and the bulge, which provides a restoring force that brings the local distribution back to the midplane. A similar mix of self-gravity and an external potential in the isothermal plane model was considered by \cite{weinberg1991} in his study of vertical modes. In \cite{darling2019} it is shown that this model admits a DMD representation. In this section, we use it to illustrate how the framework proposed in Section \ref{mapping} can be applied to a nonlinear astrophysical problem. 
	
	\subsubsection{System and Data}
	
	The equilibrium DF and corresponding density are given by
	
	\begin{equation}
		f_{\rm eq}\left (z,\,v_z\right ) = \frac{\rho_0}{\left (2\pi\sigma_z^2\right )^{1/2}} 
		e^{-E_z/\sigma_z^2},
	\end{equation}
	
	\noindent and
	
	\begin{equation}\label{eqdensity}
		\rho_{\rm eq}(z) = \rho_0 e^{-\Phi(z)/\sigma_z^2},
	\end{equation}
	
	\noindent where $ E_z $ is the vertical energy, and $ \sigma_z $ is the vertical velocity dispersion. This density, along with a corresponding equilibrium potential, comprise a potential-density pair satisfying the Poisson equation. Defining $\rho_0 = \sigma_z^2/2\pi Gz_0^2$, such a potential is 
	
	\begin{equation}\label{eqpot}
		\Phi_{\rm eq}(z) = 2\sigma_z^2 \ln{\cosh{\left (z/z_0\right )}}.
	\end{equation}
	
	\noindent In the particular system realization here, $\sigma_z = 20\,{\rm km\,s}^{-1}$ and $z_0 = 500\,{\rm pc}$, such that the surface density is $\Sigma = 2z_0\rho_0 = 60\,M_\odot\,{\rm pc}^{-2} $.  As in \cite{darling2019}, we separate the gravitational potential into an external, time-independent part, and a time-dependent part
	
	\begin{equation}\label{totalpot}
		\Phi(z,t) = (1-\alpha)\Phi_{\rm eq}(z) + \alpha\Phi_{\rm live}(z,t).
	\end{equation}
	
	\noindent The ``live fraction", $ \alpha $, quantifies the relative dominance of self-gravity and external potential. In particular, when $ \alpha \sim 1$, the system is governed by mutual interaction, and it takes very little energy to displace the system as a whole in the $ z-v_z $ plane. The dynamics in this regime are characterized by bending of the disc. Alternatively when $ \alpha \ll1$, the stars behave as test particles. So long as the potential is anharmonic, as is the case for equation \ref{eqpot}, any initial displacement of the system leads to phase space spirals of the type observed in \cite{antoja2018}. Though equation \ref{totalpot} is an abstraction, it provides a convenient means to quantify the relative importance of the disk and halo as well as the size of the patch that has been displaced from the mid-plane. 
	
	Unlike in the harmonic oscillator example, the time evolution equation (equation \ref{cbe}) is nonlinear and we cannot in general find analytic expressions for eigenfunctions of the DF. Instead, we estimate a  linear time evolution operator and its corresponding eigendecomposition with DMD. 
	
	The data for our CNN model is generated by running simulations of the isothermal plane, and then computing the eigendecomposition of the evolution operator with DMD. The simulations contain  $ N=10^5 $ stars, evolved in a simple leap-frog scheme. The orbital period of the system is set to $ T=\sqrt{2}\pi z_0/\sigma_z $, and in terms of this the time step and simulation time are $ \Delta t =T/500 $ and $ 4T $ respectively. As indicated in equation \ref{totalpot}, the forces applied to each star are the combined results of the time-dependent matter distribution given weight $ \alpha $, and the static, equilibrium potential in equation \ref{eqpot}, given weight $ 1-\alpha $. The DF is computed on a $ 60\times60 $ grid in a $ 1.4 \ \text{kpc}\times140\ \text{kms}^{-1} $ box. All simulations are initially perturbed by a $ 10 \ \text{kms}^{-1} $ displacement in velocity space.

	The inputs for the model are isolated snapshots of the simulated DFs taken at randomly generated times biased to be away from the startup period of the simulation. The outputs, which consist of the eigenfunctions and eigenvalues of the evolution operator in equation \ref{flowDMD}, are determined with DMD as described in Section \ref{datadriven}. We use the continuous time eigenvalues from equation \ref{omega} so that the model does not require knowledge of $ \Delta t $. Each simulated isothermal plane has a randomly generated $\alpha$ sampled from a probability distribution. The parameter $\alpha$ separates qualitatively different dynamics, and acts analogously to the potential curvature parameter, $\xi$, in the Schr\"{o}dinger equation example of Section \ref{qsho_section}. Like in that example, we specify a mean value, $ \bar{\alpha}=\frac{1}{2} $, and a standard deviation $ \sigma_\alpha = 0.1 $. We generate $ 1200 $ input-output pairs, with live fractions drawn from  the Gaussian distribution $N(\bar{\alpha},\sigma_\alpha)$. From this data set we use $ 1000 $ for training and $ 200 $ as a test set. For additional testing to probe how well the model generalizes we generate $ 5000 $ input-output pairs with $ \alpha $ values sampled from the uniform distribution $U(\bar{\alpha}-3\sigma_\alpha,\bar{\alpha}+3\sigma_\alpha) $. 
	
	To be clear, we do not simulate the flow of the DF $ f(q,p,t) $ generated by the CBE directly. Rather, we perform N-body simulations, evolving a finite number of discrete particles according to Hamilton's equations. It follows then that the state vectors of our system consist of the kinematic information for each of these $ N $ particles. We do not construct our data matrices directly from these vectors, but rather map them into observables first. Our observables in this case are estimates of the DF $ f(q,p,t) $, on a grid of cells in $ q $ and $ p $. Estimates of the DF at each time are computed from the number of particles per phase space cell, weighted by mass, and divided by the cell volume. The resulting matrices are then reshaped into column vectors, from which we can build a data matrix as in equation \ref{dataMatrix}.

	\subsubsection{Neural Network Model and Results}
	
	The problem of estimating $ G $ for this model is more complex than the previous example. We now have two-dimensional eigenfunctions, complex eigenvalues, and a time-dependent potential. Interestingly however the simple network architecture introduced in the previous example does not require great modification. We replace the one-dimensional configuration-space convolution operators with two-dimensional phase space ones and allow for complex eigenfunctions and eigenvalues. We reduce the number of layers so as to speed up the training process. The CNN architecture is summarized in Tables \ref{vectorTable} and \ref{valueTable}. In principle, there should be a single mode with $ \omega = 0 $, which we identify with equilibrium. However, in DMD it is common for the eigenvalue of this mode to have a small real part and zero imaginary part. The mode then slowly decays on time scales much greater than the length of the simulations. In this example, we do not estimate an imaginary component of the equilibrium eigenvalue, but rather always set it to zero. This allows the training process to focus on the complex eigenvalues of the time-dependent modes. In principle the real part may also be forced to zero for a truly stable equilibrium solution.

	We train $ G_\psi $ and $ G_\omega $ with learning rates of $ 5\times10^{-5} $, and $ 10^{-4} $, and patience parameters of $ 200 $ and $ 50 $ for the two models respectively. The networks were evaluated at the test snapshots for which we know the correct eigenvectors and eigenvalues. In this test we observe that the eigenvectors are predicted with a mean squared error of $  1.69\times10^{-5} $ which corresponds to $\sim0.3\% $ of the average eigenvector magnitude. The eigenvalue model yields a mean squared error of $ 2.92\times10^{-5} $, which is $ \sim0.02\% $ of the average eigenvalue magnitude. 
	
\begin{figure*}
		\includegraphics[width=19.9cm]{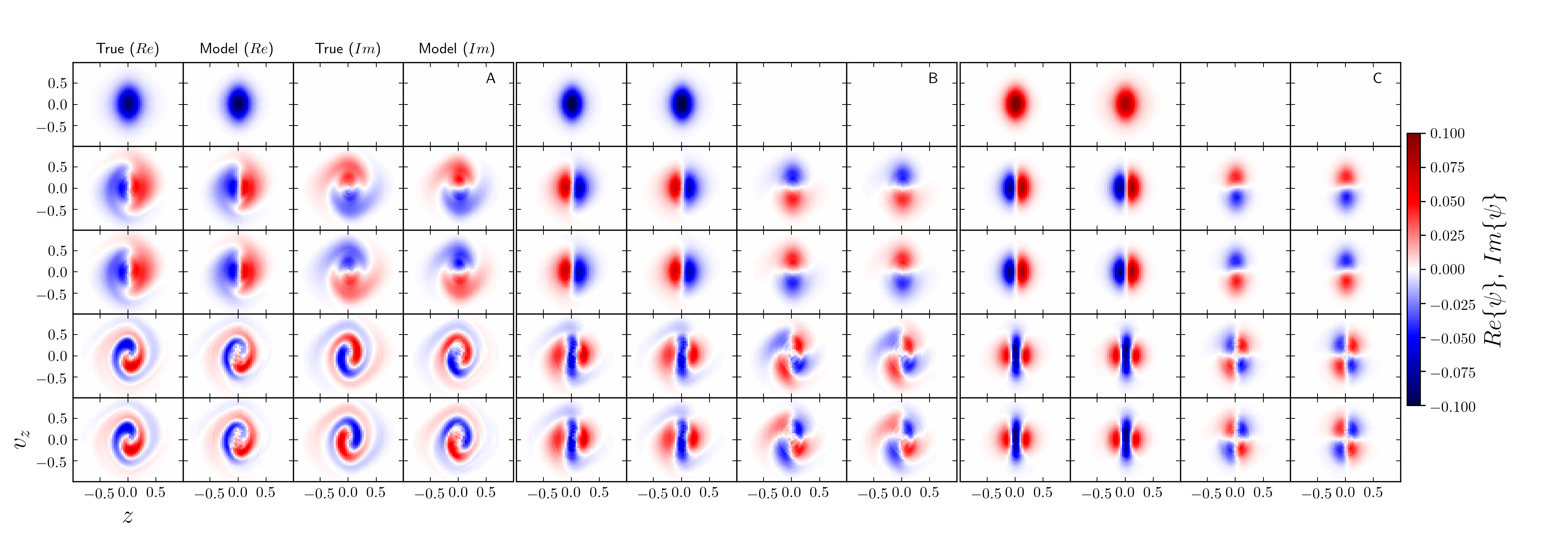}	
		\caption{Comparison of true and predicted eigenfunctions for the isothermal plane model. Row corresponds to mode number, increasing from top to bottom. The panels A, B, and C correspond to live fractions of $ \alpha=\bar{\alpha}-2\sigma_\alpha=0.3 $, $ \alpha=\bar{\alpha}=0.5 $, and $ \alpha = \bar{\alpha}+2\sigma_\alpha=0.7 $ respectively. Panels B and C follow the same layout as panel A.}
		\label{eigenvectorsTest}
	\end{figure*}

\begin{figure}
		\centering
		\includegraphics[width=7cm]{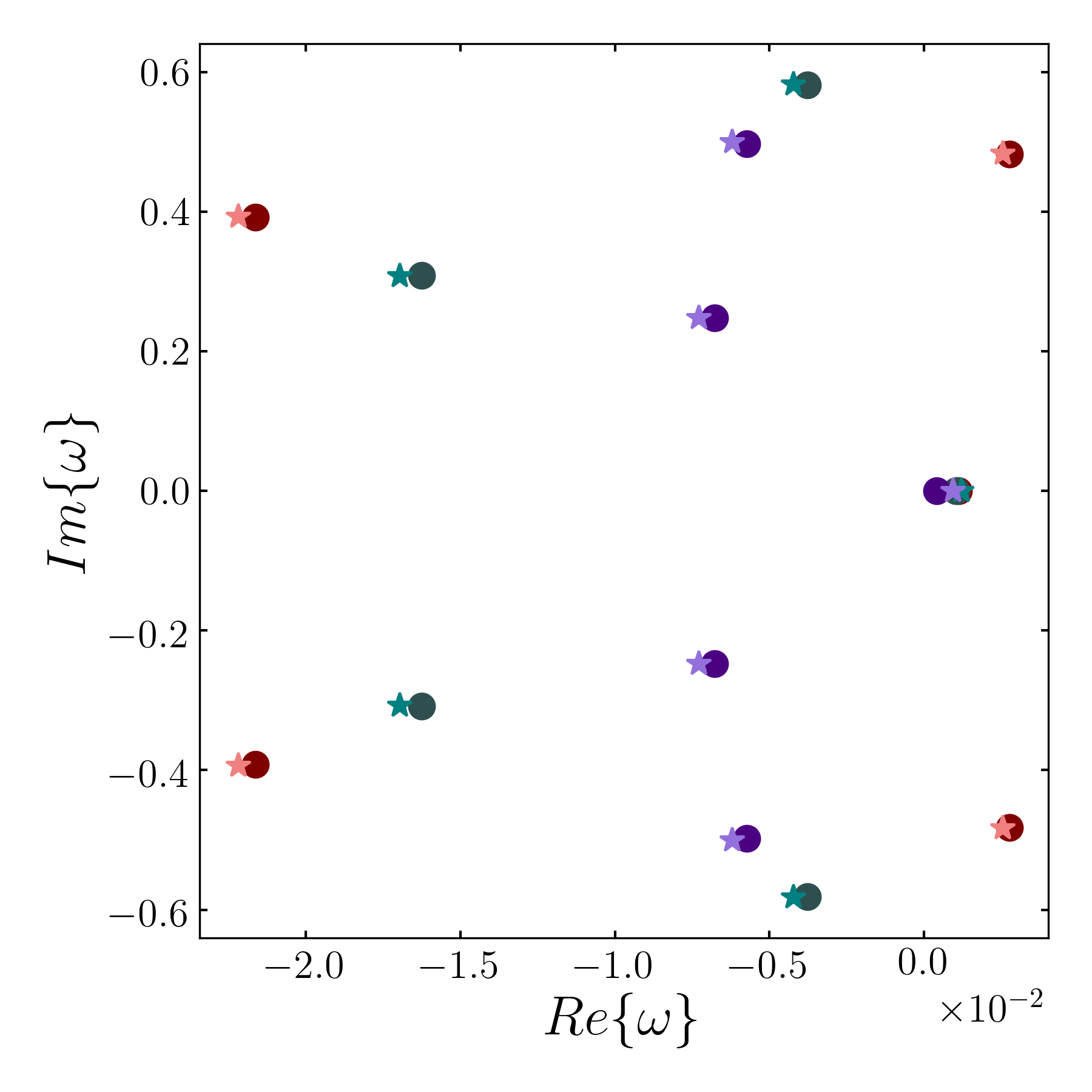}
		\caption{True and predicted eigenvalues of the evolution operator for live fractions of $ \alpha=\bar{\alpha}-2\sigma_\alpha=0.3 $ (red), $ \alpha=\bar{\alpha}=0.5 $ (green), and $ \alpha = \bar{\alpha}+2\sigma_\alpha=0.7 $ (purple). The darker colored circles indicate true values, and the lighter colored stars indicate model predictions.}
		\label{eigenvaluesIsothermal}
	\end{figure}

	Representative examples of model performance on test data are shown in Fig. \ref{eigenvectorsTest} and Fig. \ref{eigenvaluesIsothermal} for the eigenfunction and eigenvalue models respectively. Here we show test results for live fractions $ \alpha = \bar{\alpha} $, and $ \alpha=\bar{\alpha}\pm2\sigma_\alpha $. It is important to look at results on either side of the mean parameter value in this case, as the behavior of the system is drastically different for live fractions above and below $ \bar{\alpha}$. In particular, for $ \alpha $ close to zero we have kinematic particles that undergo phase mixing. For $ \alpha $ near unity the model is dominated by self-gravity, and bending and breathing modes are present \citep{weinberg1991}. The DMD modes exhibit recognizable symmetries in the $z-v_z$ plane, and one might be tempted to write each mode as a Fourier series in the phase angle coordinate of angle-action variables. Indeed, this is what is done in \citet{mathur1990} and \citet{weinberg1991} for the self-gravitating case. DMD allows for a more general and data-driven decomposition.

\begin{figure}
		\centering
		\includegraphics[width=8cm]{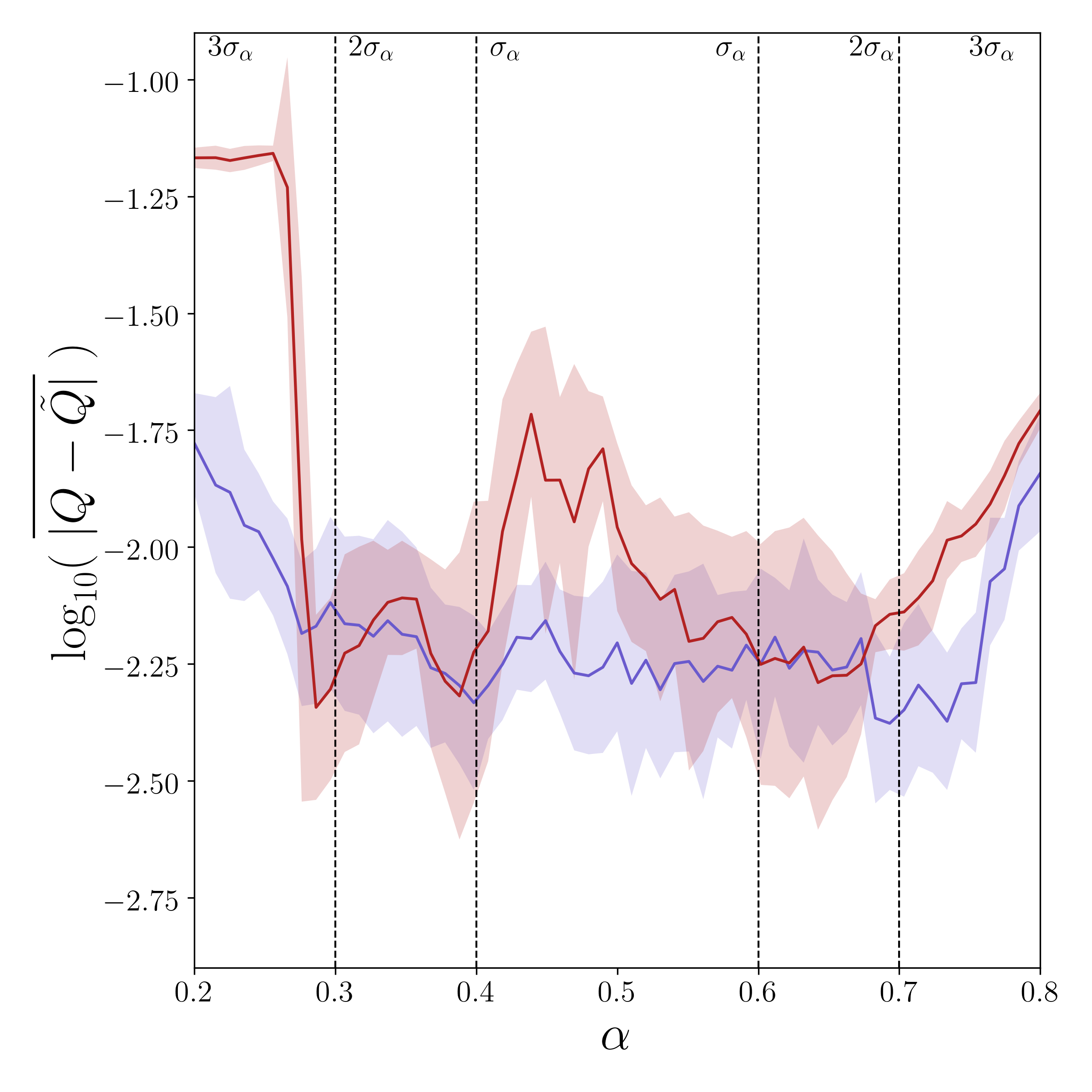}
		\caption{Logarithm of the absolute error in the quantity $ Q $ as it depends on the live-fraction $ \alpha $, with $ Q $ denoting either $ \psi $ or $ \omega $. The curves shown here are constructed by computing the average logarithmic absolute error in $ \alpha- $bins of width $ \Delta\alpha=0.1 $. The red curve shows $ Q=\omega $ and the blue curve shows $ Q=\psi $. In both cases, the band corresponds to the standard deviation of the logarithmic error within each bin.}
		\label{e_combined_iso}
	\end{figure}

	Fig. \ref{e_combined_iso} shows how the models generalize to the parameter space $ \alpha\in[\bar{\alpha}-3\sigma_\alpha,\bar{\alpha}+3\sigma_\alpha] $. Since the isothermal plane model requires N-body simulations, we are restricted in the amount of data we can generate. Therefore, we compute the average error and its standard deviation as a function of $ \alpha $ in analogy to the error densities in Fig. \ref{e_dist_vec} and \ref{e_dist_val}. The dependence of error on $ \alpha $ is then represented by a solid line for the mean value in each bin, with a transparent band representing the standard deviation. As was the case in Section \ref{qsho_section}, there is a general increase in error as one moves away from the mean, although there is a more pronounced asymmetry in the region $\alpha\in[\bar{\alpha}-\sigma_\alpha,\bar{\alpha}+\sigma_\alpha] $ for the eigenvalue predictor here than in the harmonic oscillator model, which we believe is a result of the smaller training set. Additionally like the harmonic oscillator models, the prediction errors are more tightly clustered around the maximum error for parameter values beyond $ 2\sigma_\alpha $ from the mean. This indicates that the model will consistently under-perform in this region. For relatively small training sets like those used here, one should  have sufficient knowledge of a system so as to be able to estimate the parameter to within two standard deviations of the true value. 
	
	\section{Conclusion}\label{conclusion}
	
	The challenge in galactic dynamics is to construct models that
	incorporate time-dependent phenomena but that are constrained by
	kinematic observations at a single epoch. The most promising
	approaches to this problem combine theory and simulations, each of
	which have their limitations. In particular, theoretical models
	often begin with the assumption that systems are close to
	equilibrium.  On the other hand, it can be difficult to tune
	simulations to match observations. This is because the space of possible initial
	conditions can be dauntingly large, and goodness of fit measures difficult to define.
	
	In this paper, we considered a model for the vertical structure of
	the Galactic disc, which has a single free parameter $\alpha$ quantifying the degree to which the system is self-gravitating.  When self-gravity dominates, DMD identifies bending and breathing modes of the type discussed in \citet{mathur1990, weinberg1991, widrow2014} and \citet{widrow2015}. Spiral modes are found when gravity is dominated by an external field \citep{darling2019}.  Here, we demonstrate that a CNN can recover the time-dependent DF of a system from a single snapshot of unknown $\alpha$ with the correct phase space structure and temporal characteristics. This method can be applied to the Gaia data in an effort to better understand the physics of
	the $z-v_z$ spirals found in Gaia \citep{antoja2018}.
	
	A classic example of DMD is the turbulent
	flow around an obstruction from the field of fluid mechanics. A similar situation might be at play in galactic discs where spiral structure is continually generated through a combination of swing amplification and dynamical instabilities \citep{sellwood2014,sellwood2021}. This spiral structure can then couple nonlinearly to bending of the disc, as described in \citet{MT1997a}.  The methodology presented here
	provides a novel way of exploring these phenomena. Already, preliminary results indicate that DMD can identify modes that involve both $m=2$ spirals and $m=1$ warps \citep{widrow2019}. The next step is to run a set of realistic Milky Way simulations on which a CNN can be trained to map a single snapshot to DMD modes.
	
	As with other non-parametric methods, the model is only as good as the training set. Computational time will be an issue if the method is to be applied to full three-dimensional galaxy simulations since training
	data requires computationally expensive N-body simulations. Nevertheless, we believe that DMD and CNNs are promising for efficiently extracting information about the dynamics of the Milky Way
	from the abundant incoming data from Gaia.

	\section*{Acknowledgements}
	This work was supported by a Discovery Grant with
	the Natural Sciences and Engineering Research Council of
	Canada. KD was supported by an Ontario Graduate Scholarship. We thank the reviewer for their detailed reading of our manuscript and useful suggestions. 
	
	\section*{Data Availability Statement}
	The data and software used in this article are available upon reasonable request.
	
	
	\bibliographystyle{mnras}
	\bibliography{references.bib} 
	
	\appendix

	\section{Neural Networks}
	Here we provide a very basic overview of some important details of NNs in the context of the problem described in Section \ref{mapping}. We refer the reader to \cite{goodfellow2016} for a comprehensive discussion. This section serves to provide a brief introduction for readers unfamiliar with NNs, supplying only the most necessary information on the concepts discussed in the main body of this paper.
	
	For our purposes, a NN is just a means of representing a function $ G:\mathbf{x}\mapsto\mathbf{y} $, that maps some quantity $ \mathbf{x} $ to another, $ \mathbf{y} $. In this section we will use $ \mathbf{x} $ to denote an arbitrary vector input of a NN, and $ \mathbf{X} $ an arbitrary matrix or tensor input. Additionally, we let $ \mathbf{y} $ be an arbitrary vector output, and $ \mathbf{Y} $ the matrix or tensor output. We make these distinctions here as the dimension of the input and output spaces of $ G $ depend on the dimension of the system state space. Explicitly then we have the idealized function
	
	\begin{equation}
		\mathbf{y} = G(\mathbf{x}),
	\end{equation}
	
	\noindent where the vector quantities may be exchanged for matrix or tensor quantities. We seek a representation of this function that we can use to make predictions for $ \mathbf{y} $. Consider then a particular realization
	
	\begin{equation}
		\mathbf{y}=\tilde{G}(\mathbf{x},\boldsymbol{\theta}),
	\end{equation} 
	
	\noindent where $ \boldsymbol{\theta} $ is vector of parameters that determine the particular realization of the system. We will construct NNs to be particular realizations of $ G $, which we can train to determine an optimal set of parameters $ \boldsymbol{\theta} $. The exact role the parameters play in the function depends on the architecture of the network. In what follows we give a brief overview of two basic types of architectures. The first of which we do not make explicit use of in this paper, but serves as a starting point for the second, which is used in Section \ref{examples}.
	
	\subsection{Feed Forward Neural Networks}\label{MLPbackground}
	
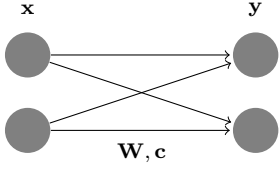
\begin{figure}\centering
		\def\layersep{1cm}
		\begin{tikzpicture}[shorten >=1pt,->,draw=black!100, node distance=\layersep]
			\tikzstyle{every pin edge}=[<-,shorten <=1pt]
			\tikzstyle{neuron}=[circle,fill=black!25,minimum size=17pt,inner sep=0pt]
			\tikzstyle{input neuron}=[neuron, fill=black!50];
			\tikzstyle{output neuron}=[neuron, fill=black!50];
			\tikzstyle{hidden neuron}=[neuron, fill=black!20];
			\tikzstyle{annot} = [text width=4em, text centered]
			
			\foreach \name / \y in {1,...,2}
			\node[input neuron] (I-\name) at (0,-\y) {};
			
			\foreach \name / \y in {1,...,2}
			\path[yshift=0.cm]
			node[output neuron] (O-\name) at (3*\layersep,-\y cm) {};
			
			\foreach \source in {1,...,2}
			\foreach \dest in {1,...,2}
			\path (I-\source) edge (O-\dest);
			
			\node[above of = I-2,  node distance=1.6cm]{$\mathbf{x}$};
			\node[above of = O-2, xshift = 0.cm, node distance=1.6cm]{\footnotesize$\mathbf{y}$};
			\node[left of = O-2, yshift = -0.3cm, node distance=1.5cm]{\footnotesize$\mathbf{W},\mathbf{c}$};
			
		\end{tikzpicture}
		\caption{Feed forward neural network with zero hidden layers, equivalent to a two dimensional linear regression problem. }
		\label{linearMLP}
	\end{figure}

\begin{figure}
		\centering
		\def\layersep{2cm}
		\begin{tikzpicture}[shorten >=1pt,->,draw=black!100, node distance=\layersep]
			\tikzstyle{every pin edge}=[<-,shorten <=1pt]
			\tikzstyle{neuron}=[circle,fill=black!25,minimum size=17pt,inner sep=0pt]
			\tikzstyle{input neuron}=[neuron, fill=black!50];
			\tikzstyle{output neuron}=[neuron, fill=black!50];
			\tikzstyle{hidden neuron}=[neuron, fill=black!20];
			\tikzstyle{annot} = [text width=4em, text centered]
			
			\foreach \name / \y in {1,...,2}
			\node[input neuron] (I-\name) at (0,-\y) {};
			
			\foreach \name / \y in {1,...,3}
			\path[yshift=0.5cm]
			node[hidden neuron] (H1-\name) at (\layersep,-\y cm) {};
			
			\foreach \name / \y in {1,...,3}
			\path[yshift=0.5cm]
			node[hidden neuron] (H2-\name) at (2*\layersep,-\y cm) {};
			
			\foreach \name / \y in {1,...,2}
			\path[yshift=0.cm]
			node[output neuron] (O-\name) at (3*\layersep,-\y cm) {};
			
			\foreach \source in {1,...,2}
			\foreach \dest in {1,...,3}
			\path (I-\source) edge (H1-\dest);
			
			\foreach \source in {1,...,3}
			\foreach \dest in {1,...,3}
			\path (H1-\source) edge (H2-\dest);
			
			\foreach \source in {1,...,3}
			\foreach \dest in {1,...,2}
			\path (H2-\source) edge (O-\dest);
			
			\node[above of = I-2,  node distance=1.6cm]{$\mathbf{x}$};
			\node[above of = O-2, xshift = 0.cm, node distance=1.6cm]{\footnotesize$\mathbf{y}$};
			\node[left of = H1-2, node distance = 1.05cm,yshift=-1.2cm]{\footnotesize$\mathbf{W}_1, \mathbf{c}_1$};
			\node[right of = H1-2, node distance = 1.05cm,yshift=-1.2cm]{\footnotesize$\mathbf{W}_2, \mathbf{c}_2$};
			\node[right of = H2-2, node distance = 1.05cm,yshift=-1.2cm]{\footnotesize$\mathbf{W}_3, \mathbf{c}_3$};
			\node[above of=H1-1, node distance=0.6cm] {\footnotesize$ \mathbf{y}_1$};
			\node[above of=H2-1, node distance=0.6cm] {\footnotesize$ \mathbf{y}_2$};
			
		\end{tikzpicture}
		\caption{Simple example of a feed forward neural network with one hidden layer. Input and output nodes are shown in black, and hidden nodes are gray.}
		\label{simpleMLP}
	\end{figure}
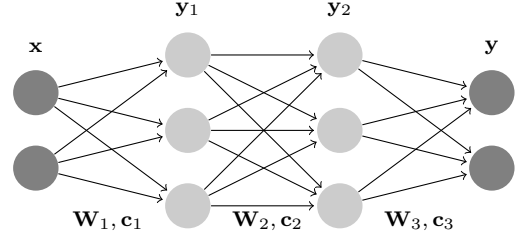

	The simplest type of NN is the feed forward network or multi-layer perceptron (MLP). At this point it is important to consider the model as segmented into layers. These consist of an input layer, output layer, and a set of hidden layers between the input and output. For MLPs, layers are comprised of parallel nodes. The number of nodes in the input and output layers must match the sizes of $ \mathbf{x} $ and $ \mathbf{y} $ respectively, but the number of nodes per hidden layer can be chosen independently. We will consider only what are called fully-connected networks here. That is, between two adjacent layers $ k $ and $ k+1 $, all nodes in layer $ k $ are connected to all nodes in layer $ k+1 $ as shown in Fig. \ref{linearMLP} or \ref{simpleMLP}. In the simplest case, the mapping from layer $ k $ to layer $ k+1 $ is linear, and is prescribed by a matrix of weights $ \mathbf{W} $ and vector of biases $ \mathbf{c} $. So for a two layer, linear network,

	\begin{equation}\label{linear}
		\mathbf{y} = \mathbf{W}^\top\mathbf{x} + \mathbf{c}.
	\end{equation}

	\noindent A graphical representation of this for the case of $ \mathbf{x}\in\mathbb{R}^2,\mathbf{y}\in\mathbb{R}^2 $ is shown in Fig. \ref{linearMLP}. Note that this is just a multivariate linear equation, and the problem of determining the optimal parameters here is the familiar linear regression problem.

	There is no reason to expect that the function we are after, $ G $ should in general be well represented by a linear function, so we would like the NN realization to be nonlinear if necessary. To introduce nonlinearity, we may use a nonlinear function $ h(x) $, called the activation function, in the mapping from layer to layer. The two layer case in equation \ref{linear} then becomes
	
	\begin{equation}\label{twolayer1}
		\mathbf{y} = h(\mathbf{W}^\top\mathbf{x} + \mathbf{c}).
	\end{equation}
	
	\noindent A commonly used nonlinear activation function is the rectified linear unit (ReLU) function given by
	
	\begin{equation}\label{twolayer2}
		\text{ReLU}(x) = \max\{0,x\}.
	\end{equation}
	
	\noindent In this paper we use only linear, $ h(x) = x $ and $ \text{ReLU} $ activation functions, opting for ReLU  at all layers except the output where we choose a linear activation function. For more on activation functions, see \cite{goodfellow2016}. 
	
	Note that in equations \ref{twolayer1} and \ref{twolayer2}, there are zero hidden layers. We now move to a simple four layer model (two hidden layers). The architecture of this network is shown in Fig. \ref{simpleMLP}, and the corresponding equations describing the mapping from input to output  are 
	
	\begin{equation}\label{4layer}
		\begin{aligned}
			\mathbf{y}_1 &= \text{ReLU}(\mathbf{W}_1^\top\mathbf{x}+\mathbf{c}_1) \\
			\mathbf{y}_2 &= \text{ReLU}(\mathbf{W}_2^\top\mathbf{y}_1+\mathbf{c}_2) \\
			\mathbf{y} &= \mathbf{W}_3^\top\mathbf{y}_2 +\mathbf{c}_3.
		\end{aligned}
	\end{equation}

	\noindent The network in Fig. \ref{simpleMLP} can also be described by table \ref{simpleMLPtable}. We will use only tables for the larger networks employed in Section \ref{examples}. 
	
	\begin{table}
		\centering
		\begin{tabular}{ c c c}\hline
			Type & Nodes & Activation \\ 
			\hline
			Dense &$ 2 $ & ReLU  \\
			Dense &$ 3 $ & ReLU  \\ 
			Dense &$ 3 $ & ReLU  \\ 
			Dense &$ 2 $ & Linear  \\ \hline
		\end{tabular}\caption{Tabular description of the network represented by Fig. \ref{simpleMLP}.}
		\label{simpleMLPtable}
	\end{table}

	The complexity of the network can be increased or decreased by adjusting the number of hidden layers, and number of nodes per layer. The mapping from input to output for an arbitrary MLP is a simple extension of equations \ref{4layer}, that is for $ N $ layers we have
	
	\begin{equation}
		\begin{aligned}
			\mathbf{y}_i &= \text{ReLU}(\mathbf{W}_i^\top\mathbf{y}_{i-1}+\mathbf{c}_i), \ \ i=1,2,...N-2, \ \ \mathbf{y}_0=\mathbf{x} \\
			\mathbf{y} &= \mathbf{W}_{N-1}^\top\mathbf{y}_{N-2} +\mathbf{c}_{N-1}.
		\end{aligned}
	\end{equation}
	
	\subsection{Convolutional Neural Networks} \label{CNNbackground}
	
	In this paper we use convolutional neural networks (CNNs). As the name implies, CNNs replace the application of a linear transformation (the weight matrix) that occurs at every layer before the application of an activation function in an MLP, with a convolution of the input matrix with a filter kernel. That is, MLP weight matrices are replaced with kernels, and the operation of linear transformation is replaced with convolution. One can think of the number of filters per layer in a CNN as analogous to the number nodes per layer in a MLP. The parameters to be learned by the network to best model the function $ G $ are the elements of the kernels. In Fig. \ref{convblock} we show a detailed view of a simple example convolutional layer.
	
	We also use what are called pooling layers. These are layers that do not have any learnable parameters, but typically apply a sampling function to their input, which computes some summary statistic of segments of the input to produce a size-reduced matrix at the output. The user specifies a ``stride'' that determines the sampling resolution. For example if one uses a stride-$ 2 $ maximum pooling layer (denoted MaxPool-2 in this paper) on a $ 60\times 60 $ input, the output is a $ 30\times 30 $ matrix where each element is the maximum of the corresponding $ 4 $-element block in the input matrix. 
	
\begin{figure}
		\centering
		
		\begin{tikzpicture}[x={(-0.7071067811865475cm,-0.7071067811865475cm)},
			y={(1cm,0cm)}, z={(0cm,1cm)}]
			\coordinate (O) at (0, 0, 0);

			\begin{scope}
				[xshift=-8,yshift=0]
				\draw[black,fill=black!50] 
				(0,0.1,0)--(0,0,0)--(0,0,2)--(0,0.1,2)--(0,0.1,0)--
				(-1.5,0.1,0)--(-1.5,0.1,2)--(0,0.1,2)--(0,0,2)--
				(-1.5,0,2)--(-1.5,0.1,2);
				
				\draw[black,fill=black!50] 
				(0,0.1,0)--(0,0.1,2)--(-1.5,0.1,2)--(-1.5,0.1,0)--(0,0.1,0);
				
				\node [black,yshift = 92, xshift=32]{$ \mathbf{X} $};
			\end{scope}
			
			\begin{scope}
				[xshift=0,yshift=0]
				\draw[black,fill=black!20] 
				(0,0.3,0)--(0,0,0)--(0,0,2)--(0,0.3,2)--(0,0.3,0)--
				(-1.5,0.3,0)--(-1.5,0.3,2)--(0,0.3,2)--(0,0,2)--
				(-1.5,0,2)--(-1.5,0.3,2);
				
				\draw[black,fill=black!20] 
				(0,0.3,0)--(0,0.3,2)--(-1.5,0.3,2)--(-1.5,0.3,0)--(0,0.3,0);
				\node [black,yshift = -6, xshift=6]{$ \mathbf{K_1} $};
			\end{scope}
			
			\begin{scope}
				[xshift=8,yshift=7]
				\draw[red,fill=red!20] 
				(-0.2,0.3,0)--(-0.2,0,0)--(-0.2,0,1.5)--(-0.2,0.3,1.5)--(-0.2,0.3,0)--
				(-1.3,0.3,0)--(-1.3,0.3,1.5)--(-0.2,0.3,1.5)--(-0.2,0,1.5)--
				(-1.3,0,1.5)--(-1.3,0.3,1.5);
				\draw[red,fill=red!20] 
				(-0.2,0.3,0)--(-1.3,0.3,0)--(-1.3,0.3,1.5)--(-0.2,0.3,1.5)--(-0.2,0.3,0);
			\end{scope}
			
			\begin{scope}
				[xshift=19,yshift=7]
				\draw[black,fill=black!20] 
				(-0.2,0.6,0)--(-0.2,0,0)--(-0.2,0,1.5)--(-0.2,0.6,1.5)--(-0.2,0.6,0)--
				(-1.3,0.6,0)--(-1.3,0.6,1.5)--(-0.2,0.6,1.5)--(-0.2,0,1.5)--
				(-1.3,0,1.5)--(-1.3,0.6,1.5);
				\draw[black,fill=black!20] 
				(-0.2,0.6,0)--(-1.3,0.6,0)--(-1.3,0.6,1.5)--(-0.2,0.6,1.5)--(-0.2,0.6,0);
				\node [black,yshift = -2, xshift=12]{$ \mathbf{K_2} $};
			\end{scope}
			
			\begin{scope}
				[xshift=39,yshift=15]
				\draw[red,fill=red!20] 
				(-0.2,0.6,0)--(-0.2,0,0)--(-0.2,0,1)--(-0.2,0.6,1)--(-0.2,0.6,0)--
				(-1.,0.6,0)--(-1.,0.6,1)--(-0.2,0.6,1)--(-0.2,0,1)--
				(-1.,0,1)--(-1.,0.6,1);
				\draw[red,fill=red!20] 
				(-0.2,0.6,0)--(-1.,0.6,0)--(-1.,0.6,1)--(-0.2,0.6,1)--(-0.2,0.6,0);
			\end{scope}
			
			\begin{scope}
				[xshift=59,yshift=15]
				\draw[black,fill=black!20] 
				(-0.2,1.2,0)--(-0.2,0,0)--(-0.2,0,1)--(-0.2,1.2,1)--(-0.2,1.2,0)--
				(-1.,1.2,0)--(-1.,1.2,1)--(-0.2,1.2,1)--(-0.2,0,1)--
				(-1.,0,1)--(-1.,1.2,1);
				\draw[black,fill=black!20] 
				(-0.2,1.2,0)--(-1.,1.2,0)--(-1.,1.2,1)--(-0.2,1.2,1)--(-0.2,1.2,0);
				\node [black,yshift = -3, xshift=20]{$ \mathbf{K_3} $};
			\end{scope}
			
			\begin{scope}
				[xshift=96,yshift=22]
				\draw[red,fill=red!20] 
				(-0.2,1.2,0)--(-0.2,0,0)--(-0.2,0,0.7)--(-0.2,1.2,0.7)--(-0.2,1.2,0)--
				(-0.7,1.2,0)--(-0.7,1.2,0.7)--(-0.2,1.2,0.7)--(-0.2,0,0.7)--
				(-0.7,0,0.7)--(-0.7,1.2,0.7);
				\draw[red,fill=red!20] 
				(-0.2,1.2,0)--(-0.7,1.2,0)--(-0.7,1.2,0.7)--(-0.2,1.2,0.7)--(-0.2,1.2,0);
			\end{scope}
			
			\begin{scope}
				[xshift=142,yshift=42]
				\draw[black,->] (0,-0.1,0) to[out=0,in=180] (0,0.3,0);
			\end{scope}
			
			\begin{scope}
				[xshift=152,yshift=12]
				\draw[black,fill=black!20]
				(0,0.1,-1)--(0,0,-1)--(0,0,3)--(0,0.1,3)--(0,0.1,-1)--
				(-0.1,0.1,-1)--(-0.1,0.1,3)--(0,0.1,3)--(0,0,3)--
				(-0.1,0,3)--(-0.1,0.1,3);
				\draw[black,fill=black!20]
				(0,0.1,-1)--(0,0.1,3)--(-0.1,0.1,3)--(-0.1,0.1,-1)--(0,0.1,-1);
				\node [black,yshift = -35, xshift=3]{\footnotesize $ \mathbf{W_1} $};
			\end{scope}

			\begin{scope}
				[xshift=162,yshift=12]
				\draw[black,fill=black!50]
				(0,0.1,-1)--(0,0,-1)--(0,0,3)--(0,0.1,3)--(0,0.1,-1)--
				(-0.1,0.1,-1)--(-0.1,0.1,3)--(0,0.1,3)--(0,0,3)--
				(-0.1,0,3)--(-0.1,0.1,3);
				\draw[black,fill=black!50]
				(0,0.1,-1)--(0,0.1,3)--(-0.1,0.1,3)--(-0.1,0.1,-1)--(0,0.1,-1);
				\node [black,yshift = 95, xshift=4]{$ \mathbf{y} $};
			\end{scope}
		\end{tikzpicture}
		
		\caption{Simple example of a NN model comprised of convolutional, pooling, and MLP layers. The input and output layers are portrayed in dark gray. Light gray layers have a ReLU activation function. MLP layers are denoted by one dimensional data volumes, and are labeled by their weight matrices $ \mathbf{W} $. Three dimensional data volumes denote convolutional and pooling layers. Convolutional layers are labeled by their filter kernels $ \mathbf{K} $, and pooling layers are shown in red.}
		\label{cnnexample}
	\end{figure}
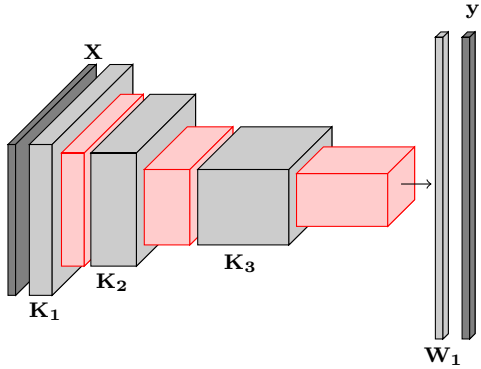

	It is also common to combine convolutional layers with the linear transformation based layers of the MLP in applications where convolutional layers are desired, but the output of the function to be modeled is not necessarily suited to the outputs of convolutional layers. An example of such a model is shown in Figure \ref{cnnexample}, and we use this overall architecture in our models. This generic structure is inspired by \cite{goodfellow2016} (see Chapter 9).
	
	\subsection{Training}
	
	To train a model, we determine the optimal set of parameters such that $ \tilde{G} $ best approximates the desired function $ G $. How to go about this is an active area of research, and we only briefly summarize the simplest training scenario to provide context to Section \ref{examples}. First we need data to train the network. To train the model, we will provide inputs $ \mathbf{x} $, and have it predict outputs $ \tilde{\mathbf{y}} $. We then use a loss function, $ L(\mathbf{y},\tilde{\mathbf{y}},\boldsymbol{\theta}) $ to quantify how well the network is doing in predicting the true $ \mathbf{y} $. Optimization of the parameters can be achieved by application of gradient descent algorithms to the loss function. In our models we use the NADAM algorithm \citep{Dozat2015IncorporatingNM}. For our purposes, the relevant quantities for training are the learning rate, which sets the step size in the gradient descent algorithm, and the patience parameter, which sets the number of iterations the optimization algorithm should tolerate with marginal change in the loss function.

	\begin{table}
		\centering
		\begin{tabular}{ c c c}
			\hline
			Type & Nodes/Filters & Activation \\ 
			\hline
			Convolution 1D& $ 8 $ & ReLU  \\ 
			Convolution 1D&$ 8 $ & ReLU \\  
			MaxPool-2 & & \\		
			Convolution 1D&$ 16 $ & ReLU  \\
			Convolution 1D&$ 16 $ & ReLU  \\
			MaxPool-2 & & \\		
			Convolution 1D&$ 32 $ & ReLU  \\
			Convolution 1D&$ 32 $ & ReLU  \\
			MaxPool-2 & & \\
			Convolution 1D&$ 16 $ & ReLU  \\
			Convolution 1D&$ 16 $ & ReLU  \\
			MaxPool-2 & & \\	
			Reshape &  & \\	
			Dense&  $ N_0 $  & Linear  \\
			\hline
		\end{tabular}\caption{Shared network architecture for both harmonic oscillator predictors. The filters consist of 9 element vectors. $ G_\psi $ uses mean squared error loss, and has output shape $ N_o=r\times M_q $. $ G_\omega $ uses mean absolute error loss and has output shape $ N_o=r $. Both models use a validation split of $ 20\% $.}\label{TableSHO}
	\end{table}

	\begin{table}
		\centering
		\begin{tabular}{ c c c}
			\hline
			Type & Nodes/Filters & Activation \\ 
			\hline
			Convolution 2D& $ 8 $ & ReLU  \\ 
			MaxPool-2 & & \\
			Convolution 2D&$ 8 $ & ReLU \\  
			MaxPool-2 & & \\		
			Convolution 2D&$ 16 $ & ReLU  \\
			MaxPool-2 & & \\		
			Convolution 2D&$ 16 $ & ReLU  \\
			MaxPool-2 & & \\	
			Reshape & & \\	
			Dense&$ r\times M_q\times M_p \times 2 $ & Linear  \\
			\hline
		\end{tabular}\caption{Network architecture of $ G_\psi $ for the isothermal plane model. The filters have  $3\times 3  $ kernels. The loss function is mean squared error, and the validation split is $ 30\% $. }\label{vectorTable}
	\end{table}
	
	\begin{table}
		\centering
		\begin{tabular}{ c c c}
			\hline
			Type & Nodes/Filters & Activation \\ 
			\hline
			Convolution 2D& $ 8 $ & ReLU  \\ 
			Convolution 2D&$ 8 $ & ReLU \\  
			MaxPool-2 & & \\		
			Convolution 2D&$ 16 $ & ReLU  \\
			Convolution 2D&$ 16 $ & ReLU  \\
			MaxPool-2 & & \\	
			Reshape & & \\	
			Dense&$ (r-1)\times 2 $ & Linear  \\
			\hline	
		\end{tabular}\caption{Network architecture of $ G_\omega $ for the isothermal plane model. The filters have  $3\times 3  $ kernels. The loss function is mean absolute error, and the validation split is $ 30\% $.}\label{valueTable}
	\end{table}
	
	\section{Software}
	
	The software used in this paper was written in the Python programming language. Packages used were: Numpy \citep{numpy}, Scipy \citep{scipy}, Tensorflow \citep{tensorflow2015}, pandas \citep{pandas} and Matplotlib \citep{matplotlib}.
	

	\bsp    
	\label{lastpage}
	
\end{document}